\documentclass[11pt]{article}

\usepackage{amsfonts}
\usepackage{graphicx}
\usepackage{amsmath}
\usepackage{amssymb}
\usepackage{latexsym}
\usepackage{color}
\usepackage{indentfirst}
\usepackage{cite}
\usepackage{float}
\usepackage{graphicx}
\usepackage{subcaption}

\begin{document}

\begin{center}
\vspace{1.8cm}


   {\Large \textbf{\ Monogamy of entanglement and steering in an atom-optomechanical system}}

\vspace{1.5cm}
\renewcommand{\thefootnote}{\alph{footnote}}
\textbf{Jamila Hmouch,}$^{1,}${\footnote{email: jamila.hmouch@edu.uiz.ac.ma}} \textbf{Mohamed Amazioug}$^{1,}${\footnote{email: {m.amazioug@uiz.ac.ma}}} \textbf{and} \textbf{Mostafa Nassik}$^{1,}${\footnote{email: m.nassik@uiz.ac.ma}}
\vspace{0.5cm}

$^{1}$\textit{LPTHE, Department of Physics, Faculty of Sciences, Ibnou Zohr University, Agadir, Morocco}\\[0.5em]

\end{center}
\section*{Abstract}
\addcontentsline{toc}{section}{abstract}
In this article, we theoretically study, in an atomic-optomechanical system, quantum correlations shared between three modes, namely mechanical mode, optical mode and atomic mode. We firstly investigate the combined effect of the variation of the cavity-collective atomic mode coupling and the effective optomechanical one, on the tripartite entanglement behavior as well as on tripartite steering evolution. Then, we discuss pairwise  entanglement and bipartite steering according to the  aforementioned couplings. Consequently, besides monogamy of the entanglement distribution, the sharing of Gaussian steering is also monogamous, thus the CKW-type monogamy inequalities are fully satisfied under all permutations of the three considered  modes in a chosen stable region. In addition, the study of tripartite entanglement and tripartite steering behaviors leads to know the optimal conditions to generate genuine tripartite entanglement, one-way and two-way steering. 
\\
\providecommand{\keywords}[1]
{
  \small	
  \textbf{\textit{Keywords---}} #1
}
 \keywords{Monogamy inequality,
             Gaussian steering,
              Logarithmic Negativity,
               Genuine tripartite entanglement}
\section{Introduction}
Several kinds of quantum correlations have been revealed, e.g, entanglement \cite{1}, 
quantum steering \cite{2,3}, Bell non-locality \cite{4} and discord\cite{5,6}; They play an important role in some of multiparty quantum mechanics  \cite{7}, such as, in computational \cite{8,9} and quantum communication tasks \cite{10,11}. Based on the important properties of Gaussian states, systems with continuous variables (CV) are the easier to manipulate experimentally than those with discrete variables\cite{12}. Focusing on optomechanics,  many studies of generation and quantification of multiparty quantum correlations using different coupling kinds, such as, optomechanical coupling \cite{13,14,15,
16,17,18,19,20,21,22}, phonon tunneling and photon hopping\cite{23, 24}.\\
On the other hand, it is well known that,  in a multiparty quantum system, the distribution of quantum correlations is limited and constrained by the monogamy property. Recalling that, monogamy is of great importance, given its potential applications in quantum information protocols, such as, entanglement distillation \cite{25}, quantum biological processes\cite{26}, quantum cryptography\cite{10}, in addition to its exploitation in characterization of quantum multiparty systems \cite{28,29, 30,31,32,33,34,35,36}. Indeed, CKW-type monogamy inequalities reflect limitations in the sharing of entanglement \cite{37}, i.e, the sum of pairwise entanglement cannot exceed the collective entanglement. Likewise, the sharing of  steering is monogamous \cite{38,39,40}, so that the sum of the degrees of steering -under Gaussian measurements- exhibited by individual pairs is less than the degree of collective steering exhibited by the Gaussian state.  Interestingly, although the CKW-type monogamy is proven for all Gaussian states \cite{41}, it is not necessarily respected by all kinds of correlation-measures \cite{42}. This is why, in what follows, two appropriate measures will be used to quantify bipartite quantum correlations, namely the contangle (square of logarithmic negativity) \cite{37} to quantify bipartite entanglement and Gaussian steering \cite{38} to quantify the degree of steering exhibited by a considered state. \\
In this paper, in Sect. \ref{2}, we introduce our atomic-optomechanical system under consideration, and we give the effective Hamiltonian. In Sect. \ref{3}, we derive and linearize the non-linear quantum Langevin equations describing the dynamics of the quadrature fluctuations, then, we develop the dynamics in the rotating wave approximation. Next, In Sect. \ref{4}, we derive the explicit formula for the covariance matrix of the three-mode state. We present in Sect. \ref{5} different quantum measures to explore quantum correlations. In Sect. \ref{6}, we discuss the numerical results. Finally, conclusions are summarized in Sect. \ref{7}.
\section{Model}\label{2} 
As shown in Fig. 1, we consider an atomic optomechanical system consisting of a Fabry-Pérot cavity composed of a fixed mirror and an oscillating end one and driven by a laser at frequency $\omega_{L}  $. The movable mirror, of a mass $ m $, has a harmonic motion which can be modeled by a mechanical mode with frequency $ \omega_{M} $ and a decay rate $ \gamma_{M} $. A set of two-level $ N_{A} $ atoms, with frequency $\omega_{A} $ and decay rate $ \gamma_{A} $, is placed inside the cavity. In what follows, the spin operators $ (S_{-},S_{+}) $ are defined as $ S_{+,-,z} = \Sigma^{ N_{A}}_{s=1} \sigma^{s}_{+,-,z} $ and satisfy the relations $[S_{+},S_{-}]= 2S_{z}$ and $[S_{z},S_{\pm}]= \pm S_{\pm}$, with, $ \sigma_{\pm} $ and $\sigma_{z} $ are the Pauli matrices. The mechanical mode position and momentum operators $ (Q$ and $P) $ satisfy $[Q,P]=$ i. $ C $ and $C^{+}$ represent the bosonic operators of annihilation and creation of the optical mode respectively, with $[C,C^{+}]= 1$. The mechanical mode $ M $ is coupled to the optical one  $ C $ by the radiation pression $G_{0}= \frac { \omega_{C}}{L} \sqrt{\frac{ \hbar}{m\omega_{M}}}$, where, $ \omega_{C} $ and $ L $ are respectively the length and the frequency of the cavity. The atom-cavity coupling constant is given by $ \Gamma_{0} =\eta\sqrt{\frac{\omega_{C}}{2\hbar\varepsilon_{0}V}} $, with $ V $ is the volume of the cavity, $ \eta $ is the dipole moment of the atomic transition and $\varepsilon_{0}$ is the vacuum permittivity. 
\begin{figure}[H]
\includegraphics[width=12cm]{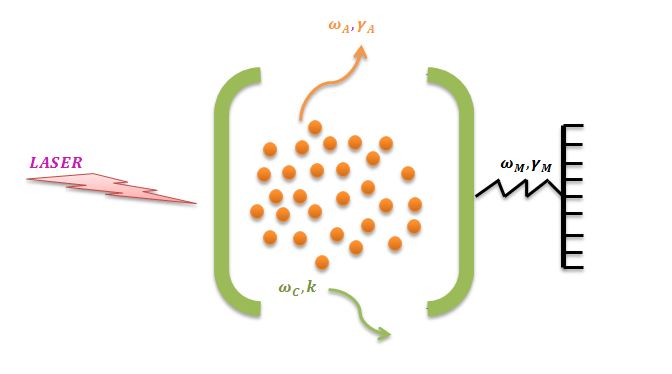}
\centering
\caption{Schematic representation of the atom-optomechanical system. The cavity is driven by a laser at frequency $\omega_{L}$. The oscillating end mirror can be modeled as a harmonic oscillator, with frequency $\omega_{M}$.}
\end{figure}
The global Hamiltonian of the system is given by ($ \hbar=1 $)
\cite{16, 43}:  
\begin{equation}
  H = \omega_{C} C^{+}C+ \frac{\omega_{A}}{2}S_{z} + \frac{\omega_{M}}{2}(Q^{2} + P^{2})+ \Gamma_{0}(S_{+}C + S_{-}C^{+}) - G_{0}C^{+}CQ + iE_{L}(C^{+}e^{-i\omega_{L}t}-Ce^{i\omega_{L}t})
\end{equation}
$ E_{L} $ describes the amplitude of the coupling between the driving laser and the cavity, which related to the input power $ P_{L} $ and the cavity decay rate $ k $ by $ \mid E_{L}\mid = \sqrt{\frac{2P_{L}k}{\hbar \omega_{L}}}$. We assume that all the atoms are initially in the ground state, so that, $S_{z}\thickapprox \langle S_{z} \rangle \thickapprox -N_{A} $, i.e, low probability of a single atomic excitation. In this case, the dynamics of the atomic polarization can be described by the bosonic annihilation operator $A=\dfrac{S_{-}}{\sqrt{\vert \langle S_{z} \rangle \vert}} $, with $[A,A^{+}]=1$ \cite{44}.
\section{Linearization of Quantum Langevin Equations}\label{3}
In the frame rotating at frequency $ \omega_{L} $, the nonlinear Langevin quantum equations describing the atomic cavity system can be written as \cite{16, 43}:
\begin{equation}
\dot{Q}= \omega_{M}P
\end{equation}
\begin{equation}
\dot{P}=-\omega_{M}Q - \gamma_{M} P + G_{0}C^{+}C +  \chi
\end{equation}
\begin{equation}
\dot{C}=-(k + i\Delta_{C})C + i G_{0} AQ - i \Gamma A + E_{L} + \sqrt{2k}C_{in}
\end{equation}
\begin{equation}
\dot{A}=-(\gamma_{A} + i\Delta_{A})C - i\Gamma A +  \sqrt{2\gamma_{A}}C_{in}
\end{equation}
with, $ \Gamma=\Gamma_{0}\sqrt{N_{A}}$, $ \Delta_{C}=\omega_{C}-\omega_{L} $($ \Delta_{A}=\omega_{A}-\omega_{L} $) is the cavity (atomic) detuning with respect to the laser. The input vacuum noise $ A_{in} $ and $ C_{in} $, with zero mean value are characterized by the non-vanishing correlations functions \cite{16}: 
\begin{equation}
\langle C^{in}(t)C^{in+}(t')\rangle=\langle A^{in}(t)A^{in+}(t')\rangle =\delta(t-t')
\end{equation}
The Langevin noise operator affecting the system (with $\langle  \chi \rangle =0$), is auto-correlated \cite{45}:
\begin{equation}
\langle   \chi(t) \chi(t')\rangle= \frac{\gamma_{M}}{2\pi\omega_{M}} \int \omega \delta(t-t') \omega[ coth(\frac{\hbar\omega}{2K_{B}T})+1]\, \mathrm{d}\omega 
\end{equation} 
where $K_{B}$ is the Boltzmann constant, $ T $ is the absolute Temperature of the mirror. The entanglement can be achieved with a very large mechanical quality factor, $ Q_{M}=\frac{\omega_{M}}{\gamma_{M}}\gg 1 $. In this limit, $  \chi $ can describe approximately a Markovian process, i.e, $  \chi$ becomes $ \delta $-correlated \cite{46}:
\begin{equation}
\langle   \chi(t) \chi(t')+   \chi(t') \chi(t)\rangle\diagup 2= \gamma_{M}(n+1)\delta(t-t')
\end{equation} 
where, $ n=[\exp (\frac{\hbar \omega_{M}}{K_{B}T})-1]^{-1} $ is the mean thermal photons number. By decomposing each operator as $ O= \langle O \rangle +  \delta O $, we solve the nonlinear quantum Langevin equations $ \eqref{2}-\eqref{5} $, so that,  $  \langle O \rangle  $ is the mean value of the operator $O$ and $  \delta O $ is its quantum fluctuation \cite{47}. The resulting stationary values can be read as: $ \langle P \rangle = 0 $; $ \langle Q \rangle = \dfrac{G_{0}\vert \langle C \rangle  \vert^{2}}{\omega_{M}} $; $ \langle C \rangle = \dfrac{E_{L}-i\Gamma \langle A \rangle }{k+i\Delta} $ and  $ \langle A \rangle =\dfrac{-i\Gamma \langle C \rangle}{\gamma_{A} + i\Delta_{A}} $. We assume that the cavity is strongly driven i.e the intracavity field has a large real amplitude ($ \vert \langle C \rangle \vert \gg 1) $, then the linear quantum Langevin equations are derived: 
\begin{equation}\label{9}
\delta \dot{Q}= \omega_{M}\delta P
\end{equation}
\begin{equation}\label{10}
\delta \dot{P}= - \omega_{M}\delta Q - \gamma_{M} \delta P + \Gamma_{0}\langle C \rangle (\delta C^{+} + \delta C) + \xi
\end{equation}
We introduce the annihilation operator of the mechanical mode:
$\delta{M}=\dfrac{\delta {Q}+ i\delta {P}}{\sqrt{2}}$, then Eqs.$ \eqref{9}-\eqref{10} $ become \cite{16, 48}:
\begin{equation}\label{11}
\delta \dot{M}= - \omega_{M}\delta M - \dfrac{ \gamma_{M}}{2}(\delta M - \delta M^{+}) + i\dfrac{G}{2} (\delta C^{+} +  \delta C)+ i\dfrac{\chi}{\sqrt{2}}
\end{equation}
\begin{equation}
\delta \dot{C}= -(k+i\Delta)\delta{C}+i \frac{G}{2}(\delta M^{+} +  \delta M)-i\Gamma \delta A + \sqrt{2k}C_{in}
\end{equation}
\begin{equation} \label{13}
\delta \dot{A}= -(\gamma_{A} + i\Delta_{A})\delta A-i\Gamma\delta C + \sqrt{2\gamma_{A}}A_{in}
\end{equation}
where $ \Delta= \Delta_{C} - \dfrac{G^{2}}{2\omega_{M}} $ 
and $ G = G_{0} \langle C \rangle \sqrt{2} $. By considering the transformations $ \delta C(t)= \delta \tilde{C}(t) e^{-i\Delta t}$, $ \delta A(t)= \delta \tilde{A}(t)e^{-i\Delta_{A}t}$, $ \delta M(t) =\tilde{M}(t).e^{-i\omega_{M}t}$, and assuming that $ \omega_{M}\gg k $,  we can neglect rapid oscillating terms $e^{\pm 2i\omega_{M}t}$ and admit that $ \Delta \thickapprox \Delta_{C} $.
In this work, we consider that $ \Delta = \Delta_{A} = -\omega_{M} $. This permits us to rewrite equations $ \eqref{11}-\eqref{13}$ in a frame rotating wave approximation ($RWA$) \cite{16} as:
\begin{equation}
\delta \dot{\tilde{M}}(t) =-\frac{ \gamma_{M}}{2}\delta \tilde{M}(t) + i \frac{G}{2} \delta \tilde{C}^{+}(t) + \sqrt{\gamma_{M}} \delta \tilde{M}^{in}
\end{equation}
\begin{equation}
\delta \dot{\tilde{C}}(t) =-k\delta \tilde{C}(t) + i \frac{G}{2} \delta \tilde{M}^{+}(t) - i\Gamma\delta \tilde{A}(t)+ \sqrt{2k} \delta \tilde{C}^{in}
\end{equation}
\begin{equation}
\delta \dot{\tilde{A}}(t) =-\gamma_{A} \delta \tilde{A}(t)-iG_{A}\delta \tilde{A}(t) + \sqrt{2\alpha} \delta \tilde{c}_{in}
\end{equation}
with $\sqrt{\gamma_{M}}M_{in}(t)= \dfrac{\omega_{M}t}{\sqrt{2}} $;  $  \tilde{C}_{in}(t)= e^{i\Delta t}C_{in} $; $  \tilde{A}_{in}(t)= e^{i\Delta t}A_{in} $. In the limit of large $\omega_{m}$, the correlation functions are:
\begin{equation}
\langle M_{in}(t)M_{in}^{+}(t')\rangle = (n+1)\delta(t-t')
\end{equation}
\begin{equation}
\langle M_{in}^{+}(t)M_{in}(t')\rangle = n\delta(t-t')
\end{equation}
Furthermore, we introduce the quadrature operators of fluctuations and noise as follows:
\begin{equation}
\delta \tilde{X}_{M}=\frac{\delta\tilde{M}^{+}+\delta\tilde{M}}{\sqrt{2}};\delta \tilde{Y}_{M}=\frac{\delta \tilde{M} -\delta \tilde{M}^{+}}{i\sqrt{2}}; \delta \tilde{X}^{in}_{M}=\frac{\delta \tilde{M}_{in}^{+}+\delta \tilde{M}_{in}}{\sqrt{2}};  \delta \tilde{Y}^{in}_{M}=\frac{\delta \tilde{M}_{in}-\delta \tilde{M}_{in}^{+}}{i\sqrt{2}}
\end{equation}
\begin{equation}
\delta \tilde{X}_{C}= \frac{\delta \tilde{C}^{+}+\delta \tilde{C}}{\sqrt{2}}; \delta \tilde{Y}_{C}=\frac{\delta \tilde{C}-\delta \tilde{C}^{+}}{i\sqrt{2}};\delta \tilde{X}^{in}_{C}=\frac{\delta \tilde{C}_{in}^{+}+\delta \tilde{C}_{in}}{\sqrt{2}};  \delta \tilde{Y}^{in}_{C}=\frac{\delta \tilde{C}_{in}-\delta \tilde{C}_{in}^{+}}{i\sqrt{2}};
\end{equation}
\begin{equation}
\delta \tilde{X}_{A}=\frac{\delta \tilde {A}^{+}+\delta \tilde{A}}{\sqrt{2}}; \delta \tilde{Y}_{A}=\frac{\delta \tilde{A}-\delta \tilde{A}^{+}}{i\sqrt{2}}; \delta \tilde{X}^{in}_{A}=\frac{\delta \tilde{A}_{in}^{+}+\delta \tilde{A}_{in}}{\sqrt{2}}; \delta \tilde{Y}^{in}_{A}=\frac{\delta \tilde{A}_{in}-\delta \tilde{A}_{in}^{+}}{i\sqrt{2}};
\end{equation}
\section{Steady state of the system}\label{4}
The steady state covariance matrix (CM) can be derived from the following equation \cite{49}:
\begin{equation}
\dot{\sqcap}(t)= \rho \sqcap(t) + \mu(t) 
\end{equation}
where, $ \sqcap^{T}(t)= (\delta \tilde{X}_{M}, \delta\tilde{Y}_{M}, \delta \tilde{X}_{C}, \delta \tilde{Y}_{C}, \delta \tilde{X}_{A}, \delta \tilde{Y}_{A}) $ is the quadrature vector,\\
$\mu^{T}(t)= (\sqrt{\gamma_{M}} \tilde{X}^{in}_{M}, \sqrt{\gamma_{M}} \tilde{Y}^{in}_{M},\sqrt{2k} \tilde{X}^{in}_{C},\sqrt{2k}\tilde{Y}^{in}_{C} ,\sqrt{2\gamma_{A}}\tilde{X}^{in}_{A},\sqrt{2\gamma_{A}} \tilde{Y}^{in}_{A})$ and $\rho$ is the drift matrix given by:
\begin{equation}
\rho = 
  \begin{bmatrix}
-\dfrac{\gamma_{M}}{2} & 0 & 0 & \dfrac{G}{2} & 0 & 0 \\
0 & -\dfrac{\gamma_{M}}{2} & \dfrac{G}{2} & 0 & 0 & 0 \\
0 & \dfrac{G}{2} &-k &0 & 0 & \Gamma \\
 \dfrac{G}{2} & 0 & 0  & -k  & -\Gamma & 0 \\
0 & 0 & 0 & \Gamma & -\gamma_{A} & 0 \\
0 & 0 & -\Gamma & 0 & 0 & -\gamma_{A} \\
\end{bmatrix} 
\end{equation}
The studied system is stable if all eigenvalues of the drift matrix $ \rho $ has negative real parts\cite{50}. The steady state of the system can be derived from the Lyapunov equation \cite{51,52}: 
\begin{equation}
\rho \Lambda + \Lambda \rho^{T}+ \Re=0
\end{equation}
where
\begin{equation}
\Re= 
\begin{bmatrix}
\dfrac{\gamma_{M}}{2}\left(2n+1\right) & 0 & 0 &0&0 & 0 \\
0 & \dfrac{\gamma_{M}}{2}\left(2n+1\right) & 0 & 0 & 0 & 0 \\
0 &0& k & 0 & 0 & 0 \\
0 & 0 & 0  & k & 0 & 0 \\
0 & 0 & 0 & 0 &\gamma_{A} & 0 \\
0 & 0 &0 & 0 & 0 &\gamma_{A} \\
\end{bmatrix} 
\end{equation}
The covariance matrix of this tripartite system can finally be written as follows:
\begin{equation}
\Lambda = 
  \begin{bmatrix}
\Lambda_{1} & 0 & 0 & \Lambda_{4} & \Lambda_{5} & 0 \\
0 & \Lambda_{1} & \Lambda_{4} & 0 & 0 & -\Lambda_{5} \\
0 & \Lambda_{4} & \Lambda_{2} & 0 & 0 & -\Lambda_{6} \\
\Lambda_{4} & 0 & 0  & \Lambda_{2} & \Lambda_{6} & 0 \\
\Lambda_{5} & 0 & 0 & \Lambda_{6} &\Lambda_{3} & 0 \\
0 & -\Lambda_{5} & -\Lambda_{6} & 0 & 0 & \Lambda_{3} \\
  \end{bmatrix}
\end{equation}  
where

$\Lambda_{1} =(16(1+2n)\Gamma^{4}(k+ \gamma_{A})\gamma_{M}+ \gamma_{A} (-G^{2}+2(k+ \gamma_{A})(2 \gamma_{A} + \gamma_{M}))(2k(1+2n)\gamma_{M})+G^{2}(2k-(1+2n)\gamma_{M}))+ 2\Gamma^{2}(G^{2}(4\gamma_{A}^{2} + 4\gamma_{A}(k+(1+n)\gamma_{M})-\gamma_{M}(-2k + (1+ 2n)\gamma_{M}))+2(1+2n)\gamma_{M}(k \gamma_{M}(2k+ \gamma_{M})+ 2 \gamma_{A}^{2}(4k + \gamma_{M})+\gamma_{A}(8 k^{2} + 4 k \gamma_{M} + \gamma_{M}^{2})))) \diagup (2(G^{2} \gamma_{A}-2( \Gamma^{2}+k \gamma_{A})\gamma_{M})(-8 \Gamma^{2}(k+ \gamma_{A})+(2k+ \gamma_{M})(G^{2}-2(k+ \gamma_{A})(2 \gamma_{A}+ \gamma_{M})))).  $ \\

$ \Lambda_{2} = (16 \Gamma^{4}(k+\gamma_{A})\gamma_{M}+ \gamma_{A}(-G^{2}+ 2(k + \gamma_{A})(2\gamma_{A} + \gamma_{M}))(2k\gamma_{M}(2k\gamma_{M}(2k+\gamma_{M})+ G^{2}(-2k + (1+ 2n)\gamma_{M}))+2 \Gamma^{2}(G^{2}(-4k\gamma_{A}-4\gamma_{A}^{2}+ \gamma_{M}(-2k+\gamma_{M}+2n\gamma_{M}))+2 \gamma_{M}(k\gamma_{M}(2k+\gamma_{M})+2\gamma_{A}^{2}(4k+\gamma_{M})+ \gamma_{A}(8k^{2}+ 4k\gamma_{M} + \gamma_{M}^{2})))) \diagup 2(G^{2}\gamma_{A}-2(\Gamma^{2}+k\gamma_{A})\gamma_{M})(-8\Gamma^{2}(k+\gamma_{A})+(2k+\gamma_{M})(G^{2}-2(k+\gamma_{A})(2\gamma_{A}+\gamma_{M})))). $\\

$ \Lambda_{3}= (16 \Gamma^{4}(k+\gamma_{A})\gamma_{M}+\gamma_{A}(2k + \gamma_{M})(G^{2}-2(k+ \gamma_{A})(2\gamma_{A}+ \gamma_{M}))+ 2\Gamma^{2}(G^{2}(-4\gamma_{A}^{2}+(1+2n)\gamma_{M}(2k + \gamma_{M})+ 4\gamma_{A}(-k + (1+n)\gamma_{M}))+ 2\gamma_{M}(k\gamma_{M}(2k +\gamma_{M}) +2\gamma_{A}^{2}(4k+\gamma_{M})+\gamma_{A}(8k^{2} + 4k\gamma_{M} +\gamma_{M}^{2}))))\diagup (2(G^{2}\gamma_{A}-2(\Gamma^{2}+ k \gamma_{A})\gamma_{M})(-8\Gamma^{2}(k + \gamma_{A})+(2k + \gamma_{M})(G^{2} - 2(k + \gamma_{A})(2\gamma_{A} + \gamma_{M})))).  $\\

$ \Lambda_{4}= 2(1+n)G\gamma_{M}(2 \Gamma^{2}(k + \gamma_{A})(2\gamma_{A} + \gamma_{M}) + k\gamma_{A}(-G^{2}+2(k +\gamma_{A})(2\gamma_{A} + \gamma_{M})))\diagup (G^{2}\gamma_{A} - 2( \Gamma^{2}+ k \gamma_{A})\gamma_{M})(-8\Gamma^{2}(k + \gamma_{A}) + (2k +\gamma_{M})(G^{2} - 2(k + \gamma_{A})(2\gamma_{A} + \gamma_{M}))). $\\

$ \Lambda_{5}= 2(1+n) \Gamma G (4(\Gamma^{2}(k + \gamma_{A})+ \gamma_{A}(G^{2} + 4k(k +\gamma_{A})))\gamma_{M} \diagup  
(G^{2}\gamma_{A}-2(\Gamma^{2}+ k \gamma_{A})\gamma_{M})(-8 \Gamma^{2}(k + \gamma_{A})+(2k +\gamma_{M})(G^{2}-2(k + \gamma_{A})(2\gamma_{A} + \gamma_{M}))). $\\

$ \Lambda_{6}= 2(1+n)\Gamma G^{2} \gamma_{A} \gamma_{M}(2k + 2\gamma_{A} + \gamma_{M})\diagup ( G^{2}\gamma_{A} -2(\gamma_{M}^{2} + k\gamma_{A})\gamma_{M})(-8\Gamma^{2}(k + \gamma_{A} )+(2k + \gamma_{M})(G^{2}-2(k + \gamma_{A})(2\gamma_{A} +\gamma_{M} ))) $.\\
This matrix should satisfy the Heisenberg-Roberson uncertainty principle $ \Lambda + $i$\dfrac{\Omega_{3}}{2} \geq 0 $  
\cite{53,54} to be a physical state, with \ \ $\Omega_{3}=\bigoplus^{3}_{s=1}$i$\sigma_{y}$ ($ \sigma_{y} $ is the y-Pauli matrix).
\section{Tripartite quantum measures}\label{5}
In this section, we will study the tripartite entanglement behavior shared in our system by the three modes: mechanical $(M)$, optical $(C)$ and atomic $(A)$, using the so-called CKW-type monogamy inequality, which is given by the following expression \cite{37}:
\begin{equation}
\varepsilon^{i\setminus jk}-\varepsilon^{i\setminus j}- \varepsilon^{i\setminus k}\geq 0  \ \ \ \   i,j,k \in \lbrace M,C,A \rbrace
\label{equ.1}
\end{equation} 
The term $ \varepsilon^{i\setminus jk} $ provides the amount of entanglement collectively shared by the mode $i$ with the remaining modes$j$ and $k$, it is quantified by the squard of one-mode versus two-mode logarithmic negativity, i.e, $ \varepsilon^{i\setminus jk} = (N^{i\setminus jk})^{2} $, where
\begin{equation}
 N^{i\setminus jk}= max[0,-ln(2 \vartheta^{i\setminus jk})]  
\end{equation}
with \ \ $ \vartheta^{i\setminus jk}= min\ \ eig\vert $i$ \Omega_{3} (U^{i\setminus jk} \sqcap^{(i,j,k)}U^{i\setminus jk})\vert$ and $\Omega_{3}=\bigoplus^{3}_{s=1}$i$\sigma_{y}$ ($ \sigma_{y} $ is the y-Pauli matrix), $ U^{i\setminus jk} $ is the matrix that inverts the sign of momentum of mode $i$ ($i\neq j \neq k$) \cite{54}.\\
The term $\varepsilon^{i\setminus j}(\varepsilon^{i\setminus k}) $ represents the quantity of entanglement between mode $i$ and mode $j(k)$, which can be measured by means of the square of one-mode versus one-mode logarithmic negativity, so that, $ \varepsilon^{i\setminus j} = (N^{i\setminus j})^{2} ( \varepsilon^{i\setminus k} = (N^{i\setminus k})^{2})$. The term $ N^{i\setminus k}$ is defined as:
\begin{equation}
  N^{i\setminus k} = max[0, -ln( 2  \vartheta^{i \vert k})]
\end{equation}
$ \vartheta^{i \vert k} $ is the smallest symplectic eigenvalue of partial transposed covariance matrix $ \sqcap_{(i,k)} =  
\begin{bmatrix}
Z_{1}(t) & Z_{3}(t)\\
Z_{3}^{T}(t) & Z_{2}(t)\\
\end{bmatrix} $, with $ \vartheta^{i \vert k} =\sqrt{\frac{\Delta -\sqrt{ \Delta^{2}-4 det \sqcap_{(i,k)}}}{2}} ;(i\neq k)$ and $ \Delta_{i\vert k} = det Z_{1}(t) 
 + det Z_{2}(t)- 2det Z_{3}(t)$.\\
We further investigate quantitatively the behavior of the Gaussian steering by using the following CKW-type monogamy -of Gaussian steering- inequalities\cite{38}:
\begin{equation}
G^{k \rightarrow  ij}-G^{k \rightarrow  i}- G^{k \rightarrow j}\geq 0 \ \ \ \ i,j,k \in \lbrace M,C,A \rbrace
\label{equ.1}
\end{equation} 
\begin{equation}
G^{ij \rightarrow  k }-G^{i \rightarrow  k}- G^{j \rightarrow  k}\geq 0
\label{}
\end{equation} 
where $G^{ij \rightarrow  k }(G^{k \rightarrow  ij})$ is the degree of simultaneous Gaussian steering between one mode (two modes) and the other two remaining modes (the third remaining mode), and $G^{n \rightarrow l}(n,l \in \lbrace i,j,k \rbrace ) $ is the degree of Gaussian steering between the individual pairs.\\
For a bipartite-mode Gaussian state ($ n_{
\Sigma_{1}}+n_{
\Sigma_{2}}$) \cite{38}
\begin{equation}
G^{\Sigma_{1} \rightarrow  \Sigma_{2} }:= max \lbrace 0,-\Sigma^{}_{s=1: \theta_{s}<1} ln(\theta_{s}) \rbrace
\label{}
\end{equation} with $ \Sigma =\begin{bmatrix}
\Sigma_{1}(t) & \Sigma_{3}(t)\\
\Sigma_{3}(t)^{T} & \Sigma_{2}(t)\\
\end{bmatrix} $, where $ \theta_{s} $ are the symplectic eigenvalues of the matrix $ \Omega= 
\Sigma_{2}-
\Sigma_{3}^{T}
\Sigma_{1}
\Sigma_{3} $, $ \Omega $ is the Schur complement of $ \Sigma_{1}$ in $ \Sigma $. 
\section{Analyse and Discussion}\label{6}
By using the experimental parameters reported in \cite{16, 55} $\lbrace\frac{\omega_{M}}{2\pi}= 947\times10^{3}$Hz, the mechanical damping rate $\frac{\gamma_{M}}{2\pi}=140$Hz and the mass of the movable mirror $m= 145$ng, the cavity frequency is $\frac{\omega_{C}}{2\pi}= 5.26\times10^{14}$Hz, the laser frequency $\frac{\omega_{L}}{2\pi}=2.82\times10^{14}$Hz$\rbrace$, we study the behaviors of both tripartite entanglement and tripartite Gaussian steering according to the coupling between the cavity and the collective atomic mode ($\Gamma$) and the effective optomechanical coupling $G$.
We plot the quantities $\lbrace \varepsilon^{i\setminus jk}-\varepsilon^{i\setminus j}- \varepsilon^{i\setminus k}$, $G^{k\rightarrow  ij}-G^{k\rightarrow  i}- G^{k\rightarrow  j} $ and $ G^{ij\rightarrow  k }-G^{i \rightarrow  k}- G^{j\rightarrow  k} \rbrace$, 
$i,j,k \in \lbrace M,C,A \rbrace $, in a chosen stable region which corresponds to $ 50 \times10^{3}$Hz$ \leq \Gamma \leq 54 \times10^{3}$Hz  and $20\times 10^{3}$Hz$ \leq G \leq 20.5 \times10^{3} $Hz. Moreover, we plot the quantities $\lbrace\varepsilon^{i\setminus jk}, \varepsilon^{i\setminus j}, \varepsilon^{i\setminus k}, G^{k \rightarrow  ij}, G^{ij \rightarrow  k } $, $G^{i \rightarrow  k} $ and $ G^{k \rightarrow  i} \rbrace$, 
$i,j,k \in \lbrace M,C,A \rbrace $ to provide more details.
\subsection{Tripartite entanglement}
\begin{figure}[H]
\begin{subfigure}[b]{.48\linewidth}
\fbox{\includegraphics[width=\linewidth]{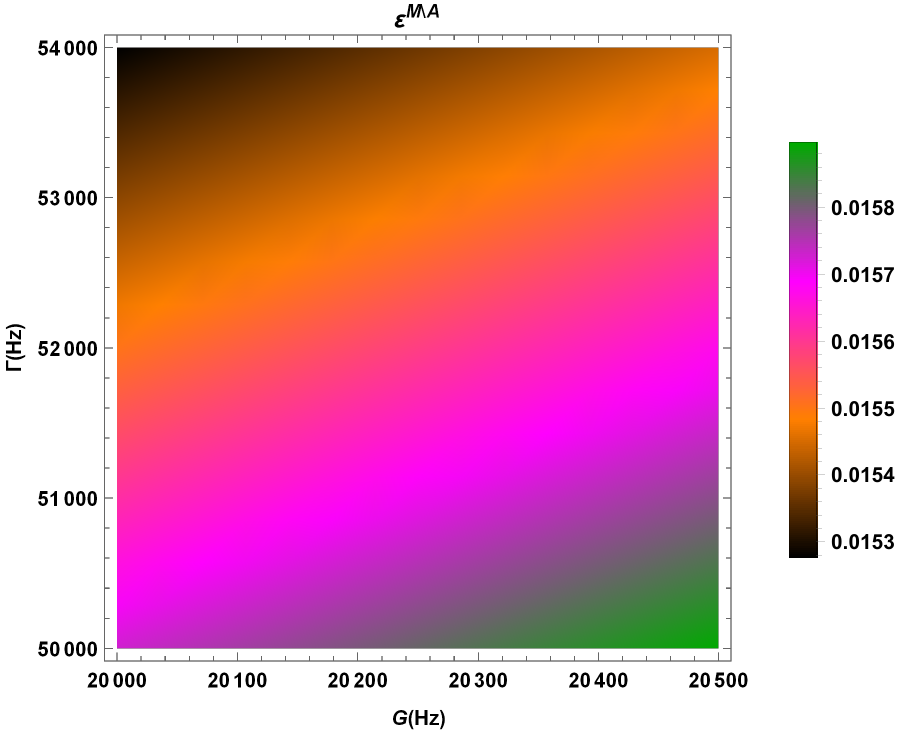}}
\caption{The amount of pairwise entanglement increases by increasing the values of $G$ and decreasing the $\Gamma$ values}\label{}
\end{subfigure}
\begin{subfigure}[b]{.48\linewidth}
\fbox{\includegraphics[width=\linewidth]{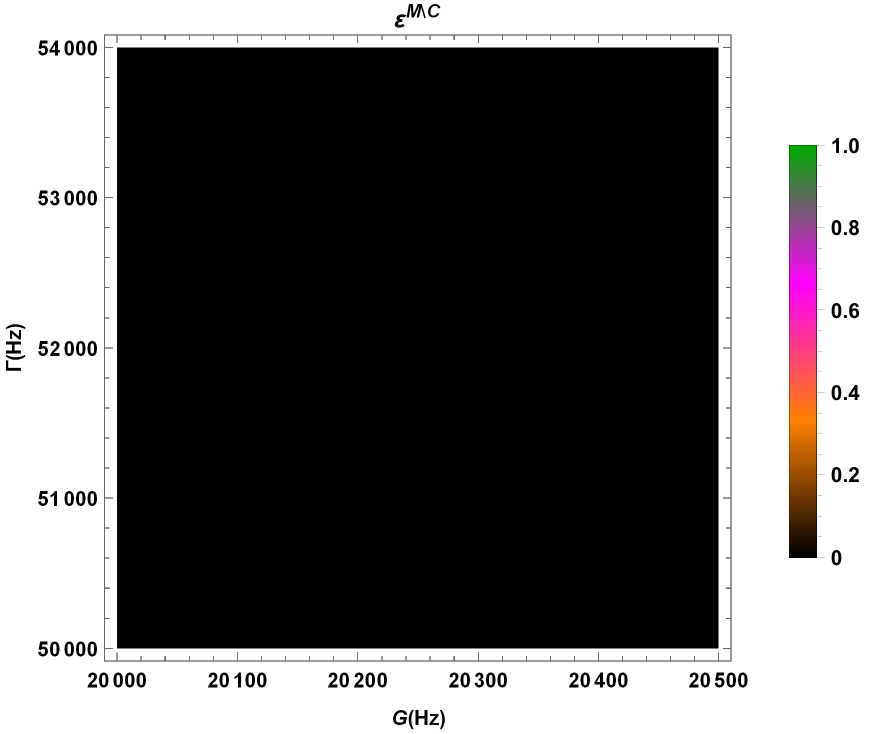}}
\caption{The entanglement between mode $M$ and mode $C$ remains zero despite the  variation of $G$ and $\Gamma$ values.}\label{}
\end{subfigure}
\begin{subfigure}[b]{.48\linewidth}
\fbox{\includegraphics[width=\linewidth]{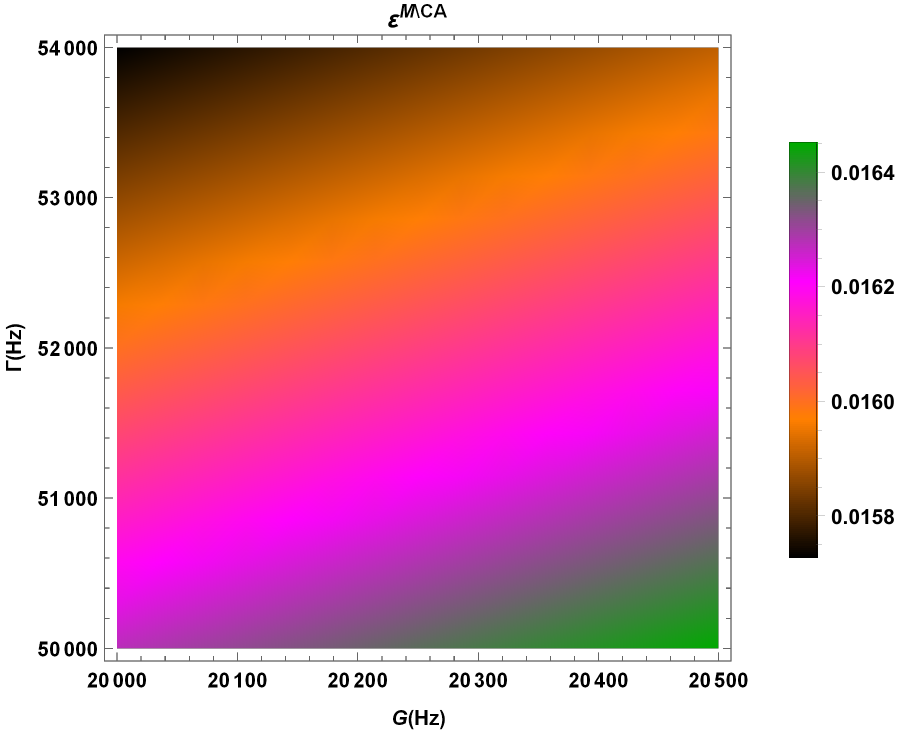}}
\caption{The behavior of collective entanglement $\varepsilon^{M\setminus CA}$ is the same as that of individual entanglement $\varepsilon^{M\setminus A}$}\label{}
\end{subfigure}
\hspace{0.5cm}
\begin{subfigure}[b]{.48\linewidth}
\fbox{\includegraphics[width=\linewidth]{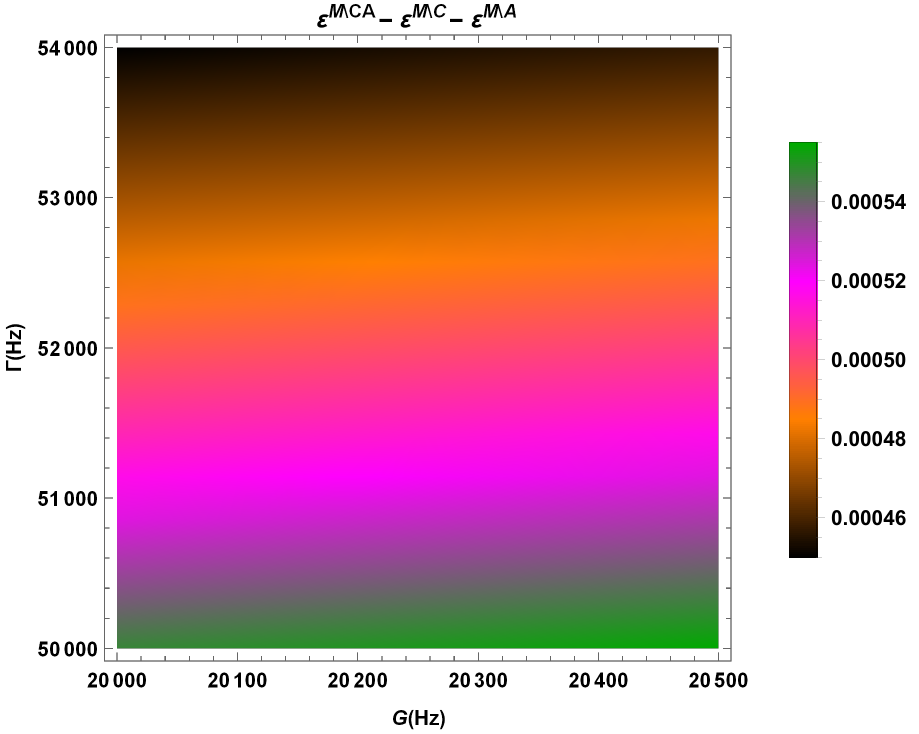}}
\caption{For any value of $G$ and $ \Gamma $, $ \varepsilon^{M\setminus CA}  \geq  \varepsilon^{M\setminus C } + \varepsilon^{M\setminus A} $ is fully satisfied }\label{}
\end{subfigure}
\caption{Bipartite entanglement $ \varepsilon^{M\setminus A} (a)$, $ \varepsilon^{M\setminus C}(b)$ and $ \varepsilon^{M\setminus CA}(c) $ behaviors versus $ \Gamma $ and $G $; $(d)$: Effect of $ \Gamma $ and $G $ on tripartite entanglement $ \varepsilon^{M \setminus CA} - \varepsilon^{M \setminus C} - \varepsilon^{M \setminus A} $, for $L= 1$ mm, $ \gamma_{A}=2\pi\times10^{3}$Hz, $k=2\pi\times10^{3}$Hz and $T=1\mu$k.}
\end{figure}
In the configurations $ \varepsilon^{M\setminus CA} $, $ \varepsilon^{M\setminus C} $ and $ \varepsilon^{M\setminus A} $, the monogamy inequality is satisfied, i.e, $ \varepsilon^{M\setminus CA}  \geq  \varepsilon^{M\setminus C } + \varepsilon^{M\setminus A} $. As shown in Figures $2(a)$, $2(b)$ and $2(c)$, this property is  due to the fact that in any region delimited by a chosen range of $ \Gamma$ and $G$ values, the quantity of entanglement shared between the mechanical mode $(M)$ and the atomic mode $(A)$ is sufficiently strong that it approaches shared collective entanglement by the mechanical mode with the optical $(C)$ and atomic $(A)$ modes; this implies that there is no individual entanglement sharing between the mechanical mode and the optical one. As a result, the sum of entanglement between each pairs remains less than the collective distributed entanglement. The variation in $G$ has a negligible effect on the monogamy inequality, while the decrease in the $\Gamma$ values slightly enhances the verification of this inequality.\\ 
\begin{figure}[H]
\begin{subfigure}[b]{.48\linewidth}
\fbox{\includegraphics[width=\linewidth]{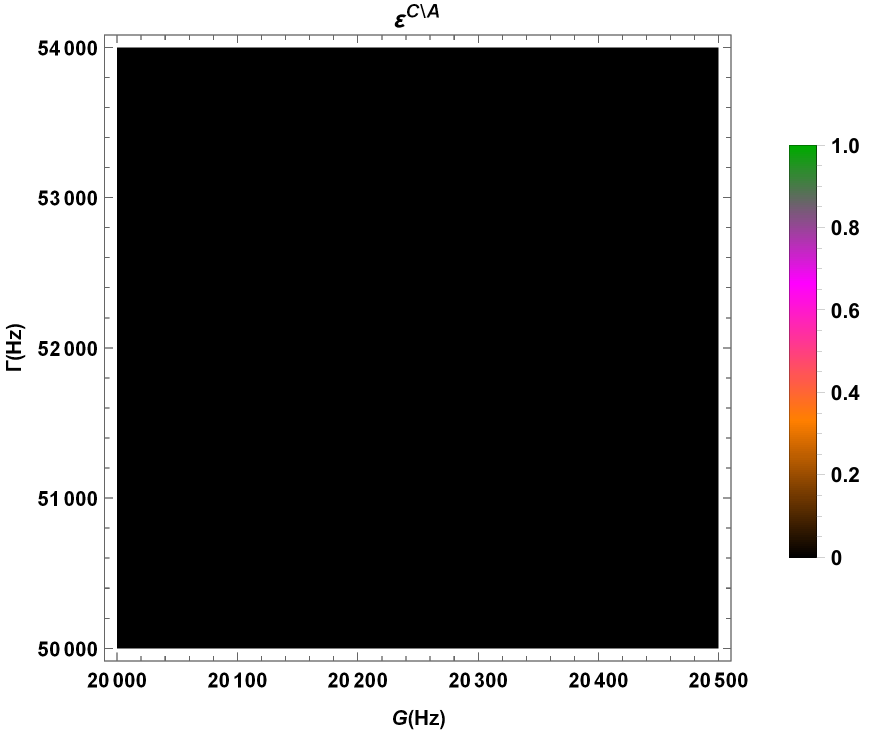}}
\caption{The entanglement between mode $C$ and mode $A$ remains zero despite the variation of $G$ and $\Gamma$ values.}\label{}
\end{subfigure}
\begin{subfigure}[b]{.48\linewidth}
\fbox{\includegraphics[width=\linewidth]{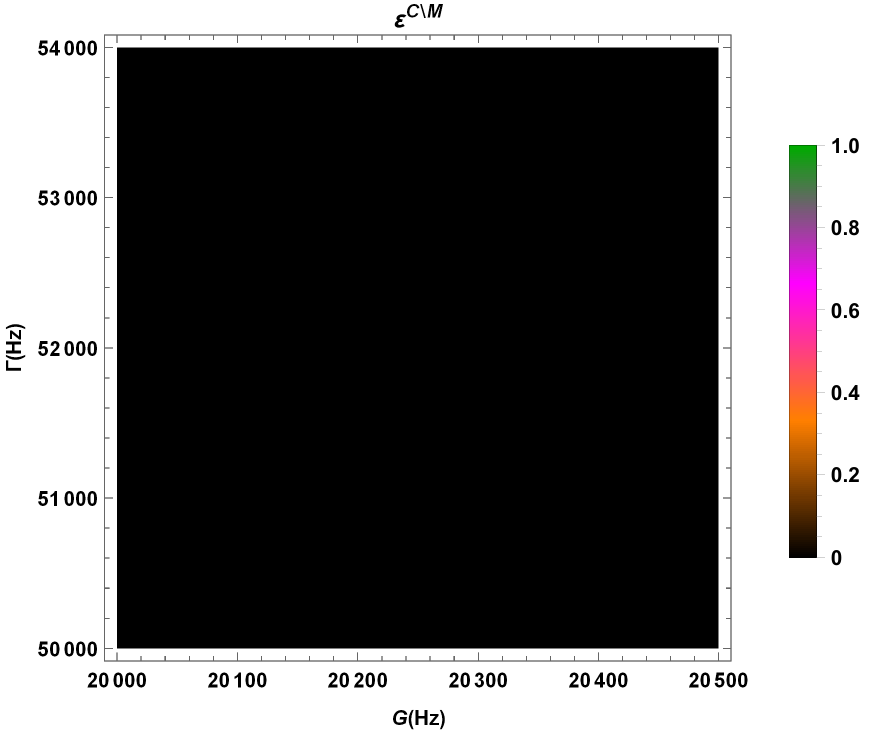}}
\caption{The entanglement between mode $C$ and mode $M$ remains zero despite the variation of $G$ and $\Gamma$ values.}\label{}
\end{subfigure}
\begin{subfigure}[b]{.48\linewidth}
\fbox{\includegraphics[width=\linewidth]{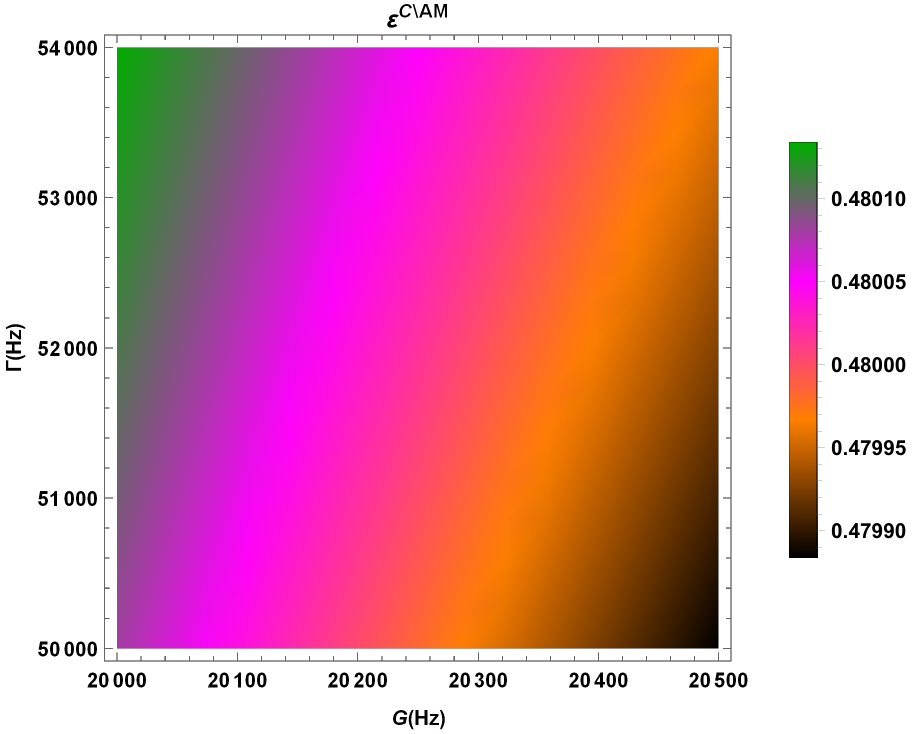}}
\caption{For higher values of $\Gamma$ and lowest values of $G$  the collective entanglement $\varepsilon^{C\setminus AM}$ becomes important}\label{}
\end{subfigure}
\hspace{0.5cm}
\begin{subfigure}[b]{.48\linewidth}
\fbox{\includegraphics[width=\linewidth]{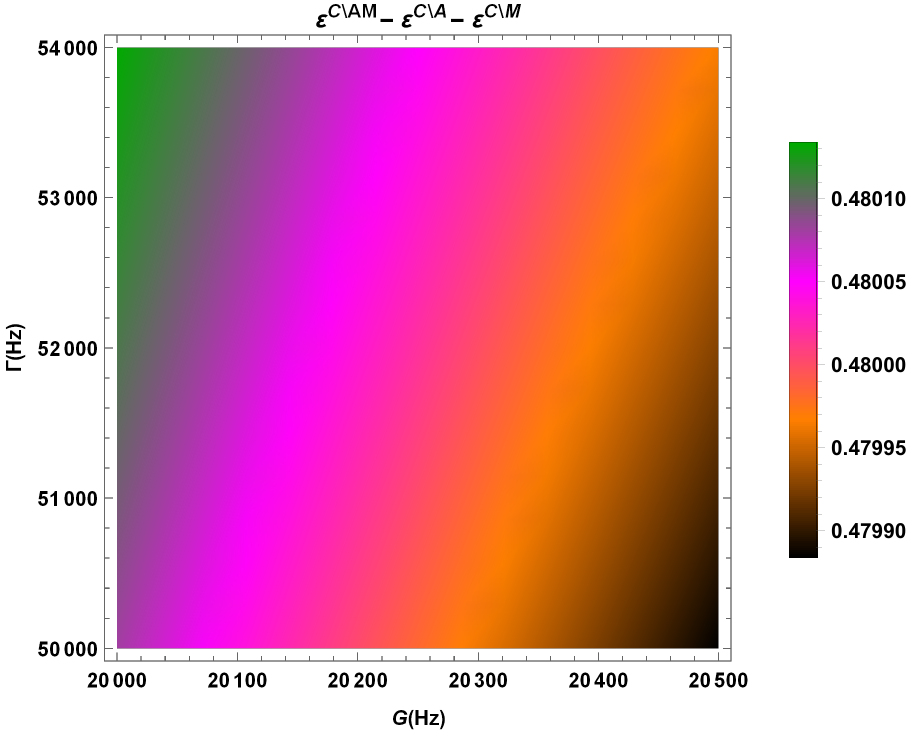}}
\caption{The evolution of $ \varepsilon^{C\setminus AM} - \varepsilon^{C\setminus A} - \varepsilon^{C\setminus M} $ is similar to that of $ \varepsilon^{C\setminus AM} $}\label{}
\end{subfigure}
\caption{Bipartite entanglement $ \varepsilon^{C\setminus A} (a)$, $ \varepsilon^{C\setminus M}(b)$ and $ \varepsilon^{C\setminus AM}(c) $ behaviors versus $ \Gamma $ and $G $; $(d)$: Effect of $ \Gamma $ and $G $ on tripartite entanglement $ \varepsilon^{C \setminus AM} - \varepsilon^{C \setminus A} - \varepsilon^{C \setminus M} $, for $L= 1$mm , $ \gamma_{A}=2\pi\times10^{3}$Hz, $k=2\pi\times10^{3}$Hz and $T=1\mu$k.}
\end{figure}
According to Fig. $3(d)$, entanglement monogamy inequality  $ \varepsilon^{C\setminus AM} - \varepsilon^{C\setminus A } - \varepsilon^{C\setminus M} \geq 0 $ holds in the area limited by the whole studied ranges of $G$ and $\Gamma$ values.
In the absence of individual sharing entanglement between optical mode and the other remaining modes, monogamy inequality in configurations $ \varepsilon^{C\setminus AM} $, $ \varepsilon^{C\setminus A} $ and $ \varepsilon^{C \setminus M} $, can be reduced to $ \varepsilon^{C\setminus AM} \geq 0 $, which, therefore, explains the similarity between the $\varepsilon^{C \setminus AM} - \varepsilon^{C \setminus A} - \varepsilon^{C \setminus M}   $ and $\varepsilon^{C\setminus AM}$ behaviors, so that, by decreasing the values of $G$ and -in parallel- increasing the values of $\Gamma$, the amount of $\varepsilon^{C\setminus AM}$ increases slowly.\\
\begin{figure}[H]
\begin{subfigure}[b]{.48\linewidth}
\fbox{\includegraphics[width=\linewidth]{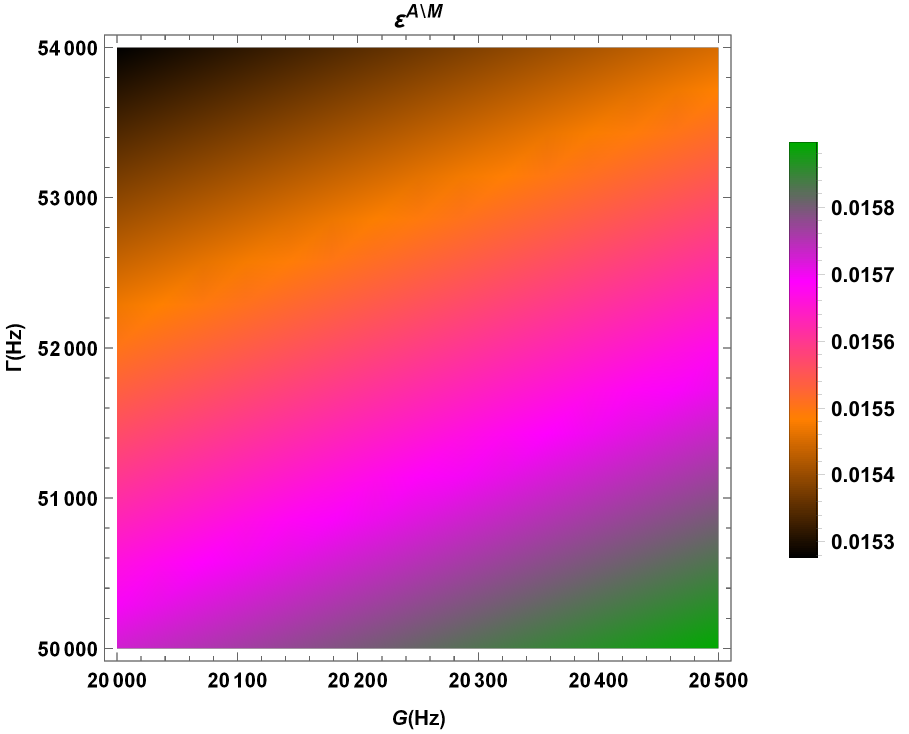}}
\caption{The amount of pairwise entanglement $ \varepsilon^{A\setminus M}$ increases by increasing the values of $G$ and decreasing the $\Gamma$ values.}\label{}
\end{subfigure}
\begin{subfigure}[b]{.48\linewidth}
\fbox{\includegraphics[width=\linewidth]{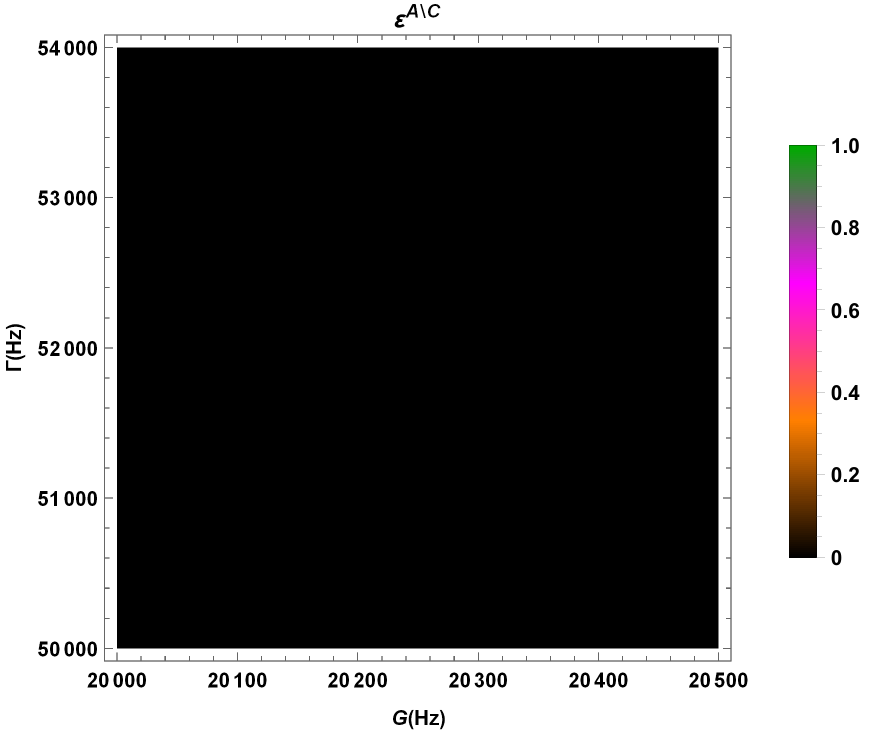}}
\caption{The entanglement between mode $A$ and mode $C$ remains zero despite the variation of $G$ and $\Gamma$ values.\ \ \ \    }\label{}
\end{subfigure}
\begin{subfigure}[b]{.48\linewidth}
\fbox{\includegraphics[width=\linewidth]{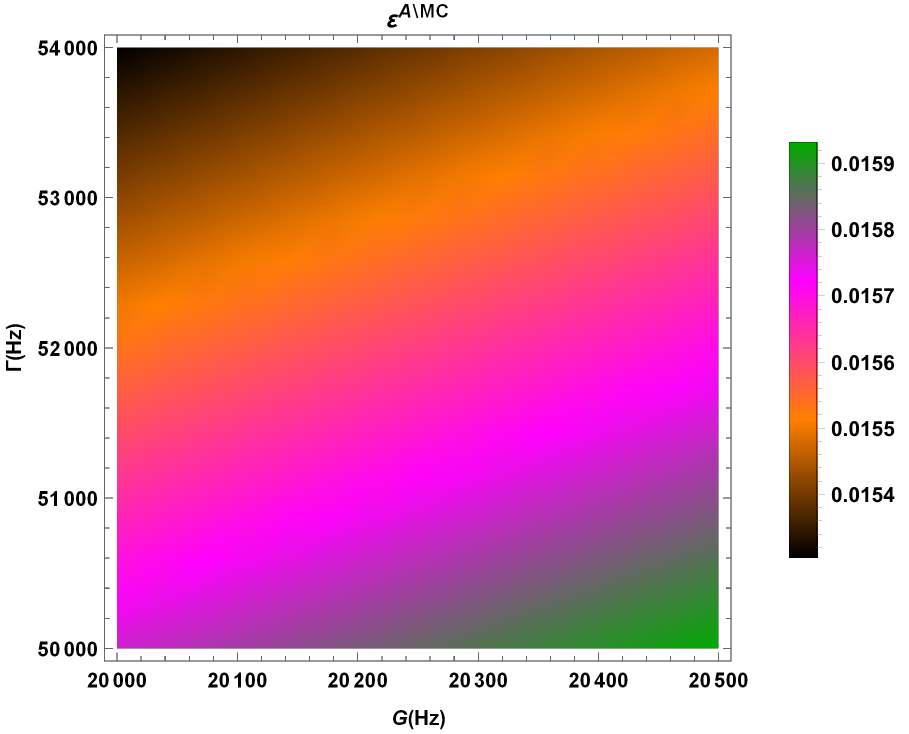}}
\caption{The amount of collective entanglement 
$ \varepsilon^{A\setminus MC} $ increases by increasing the values of $G$ and decreasing the $\Gamma$ values}\label{}
\end{subfigure}
\hspace{0.5cm}
\begin{subfigure}[b]{.48\linewidth}
\fbox{\includegraphics[width=\linewidth]{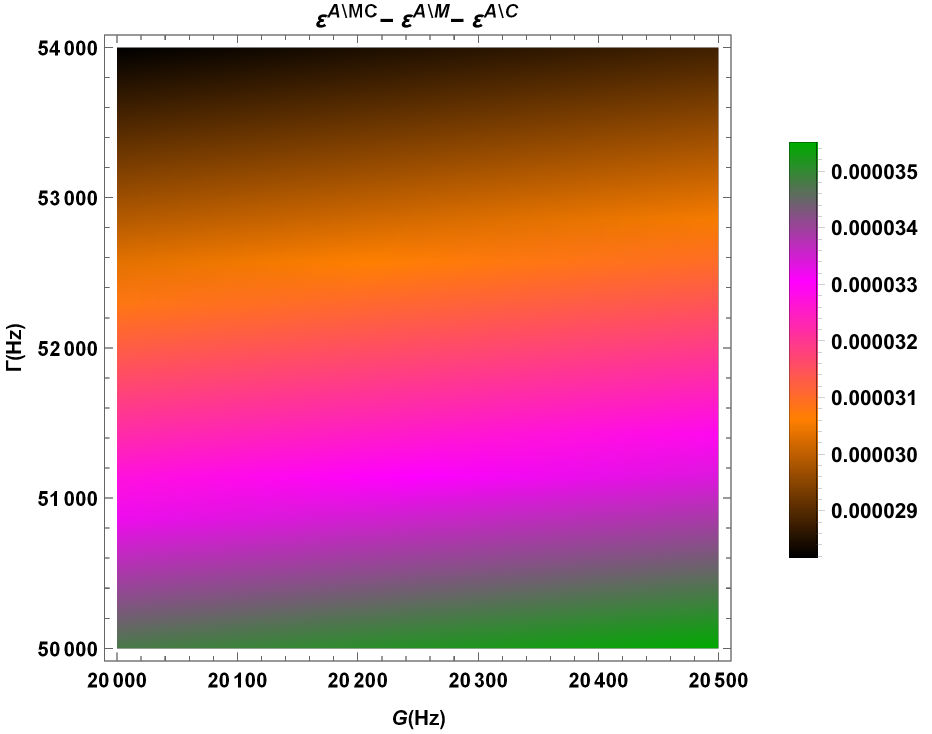}}
\caption{For any value of $G$ and $ \Gamma $, the inequality $ \varepsilon^{A\setminus MC} - \varepsilon^{A\setminus M } - \varepsilon^{A\setminus C} \geq 0 $ is fully satisfied. \ \ \ \ \ \ \ \ \ \ \ \ \ \ \ \ \ \ \ \ \ \ \ \ \ \ \ \ \ \ \ \ \ \ \ \ \ \ \ \ \ \ \ \ \ \ \ \ \ \ \ \ \ \ \ \ \ \ \ \ \ \ \ \ \ \ \ \ \ \ \ \ \ \ \ \ \ \ \ \ \ \ \ \ \ \ \ \ \ \ \ \ \ \ \ \ }\label{}
\end{subfigure}
\caption{Bipartite entanglement $ \varepsilon^{A\setminus M} (a)$, $ \varepsilon^{A\setminus C}(b)$ and $ \varepsilon^{A\setminus MC}(c) $ behaviors versus $ \Gamma $ and $G $, $(d)$: Effect of $ \Gamma $ and $G $ on tripartite entanglement $ \varepsilon^{A \setminus MC} - \varepsilon^{A \setminus M} -  \varepsilon^{A \setminus C}$, for $L= 1$ mm, $ \gamma_{A}=2\pi\times10^{3}$Hz, $k=2\pi\times10^{3}$Hz and $T=1\mu$k.}
\end{figure}
The verification of entanglement inequality can be observed in configurations $ \varepsilon^{A\setminus MC} $, $ \varepsilon^{A\setminus M} $ and $ \varepsilon^{A \setminus C} $,  i.e, $ \varepsilon^{A\setminus MC} - \varepsilon^{A\setminus M}-\varepsilon^{A\setminus C} \geq 0  $. The similar behavior of entanglement monogamy in Figures $4(d)$ and $2(d)$ implies the same effect of varying the values of parameters  $\Gamma$ and $G$ on entanglement monogamy relationship. Based on Fig. $4(a)$, $4(b)$ and $4(c)$, it is evident that the entanglement monogamy relationship evolution depends only on the variation of the quantity of entanglement  collectively shared and the individual entanglement amount  between mode $A $ and mode $M$.\\
Concluding this section, we can say that the inequalities of entanglement monogamy are fully satisfied under all permutations of the three modes considered, which testifies that the entanglement distributed between them is monogamous. Moreover, Fig. $2(d)$ shows that the studied three-mode Gaussian state exhibits a genuine tripartite entanglement quantified by $\varepsilon^{A \setminus MC} - \varepsilon^{A \setminus M} - \varepsilon^{A \setminus C}$ which represents the minimum degree of monogamy compared to the quantities $\varepsilon^{C \setminus AM } - \varepsilon^{C \setminus A } - \varepsilon^{C \setminus M }   $ and $\varepsilon^{M \setminus CA} - \varepsilon^{M \setminus C} - \varepsilon^{M \setminus A} $. According to Fig. 4(d), genuine tripartite entanglement can be displayed by relaxing the atomic-cavity coupling $\Gamma$ for a sufficiently large value of $G$. 

\subsection{Tripartite steering}
\begin{figure}[H]
\begin{subfigure}[b]{.48\linewidth}
\fbox{\includegraphics[width=\linewidth]{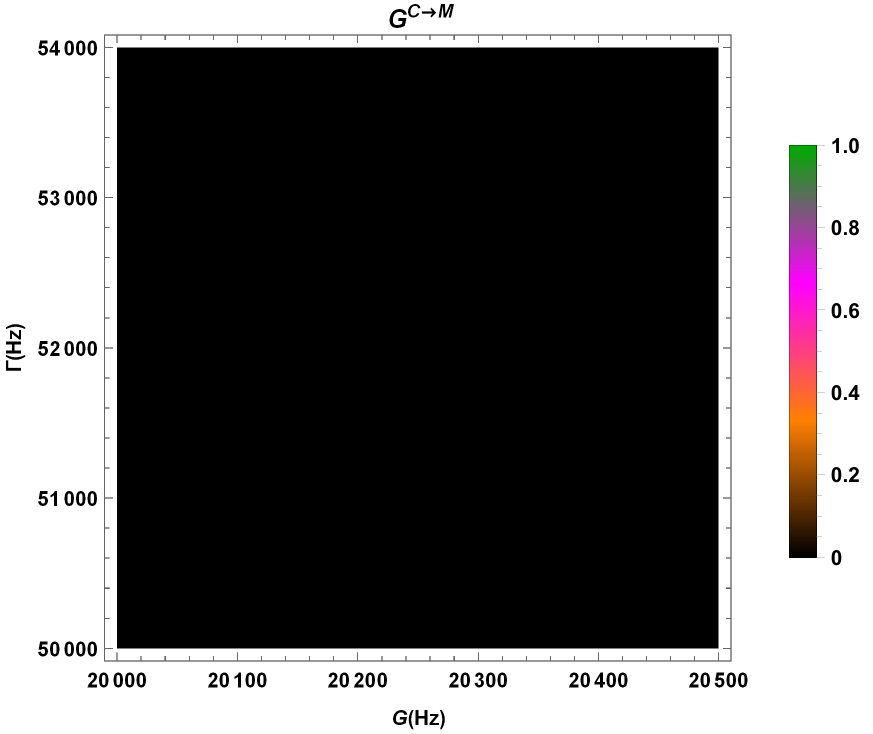}}
\caption{The steering of the mode $(M)$ by the mode $(C)$ is not authorized.}\label{}
\end{subfigure}
\begin{subfigure}[b]{.48\linewidth}
\fbox{\includegraphics[width=\linewidth]{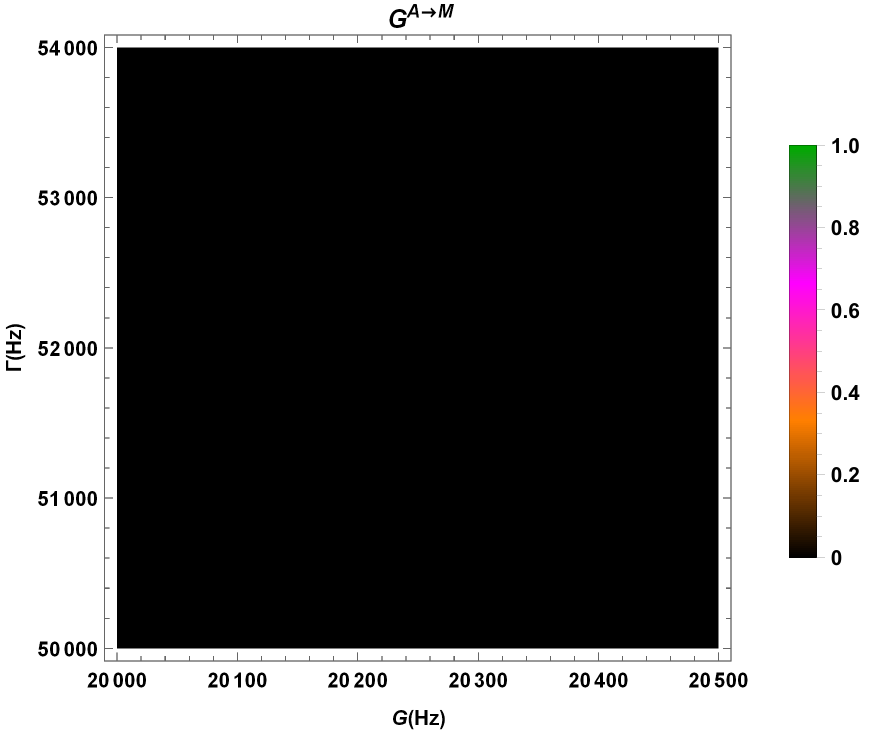}}
\caption{The steering of the mode $(M)$ by the mode $(A)$ is not authorized.    }\label{}
\end{subfigure}
\begin{subfigure}[b]{.48\linewidth}
\fbox{\includegraphics[width=\linewidth]{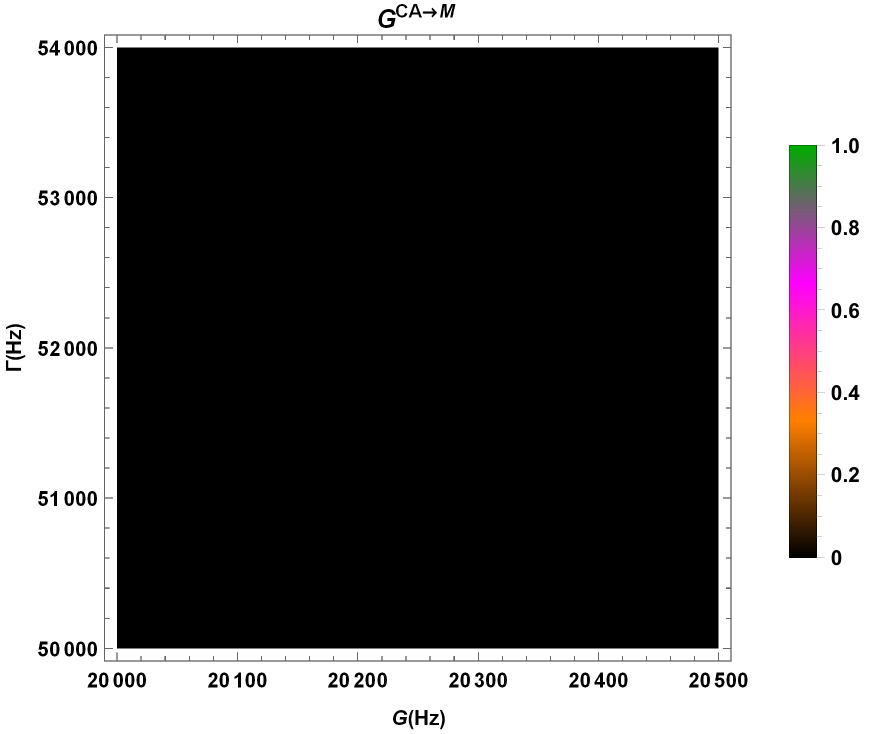}}
\caption{No collective steering of the mode $M$ by the modes $C$ and $A$ }\label{}
\end{subfigure}
\hspace{0.5cm}
\begin{subfigure}[b]{.48\linewidth}
\fbox{\includegraphics[width=\linewidth]{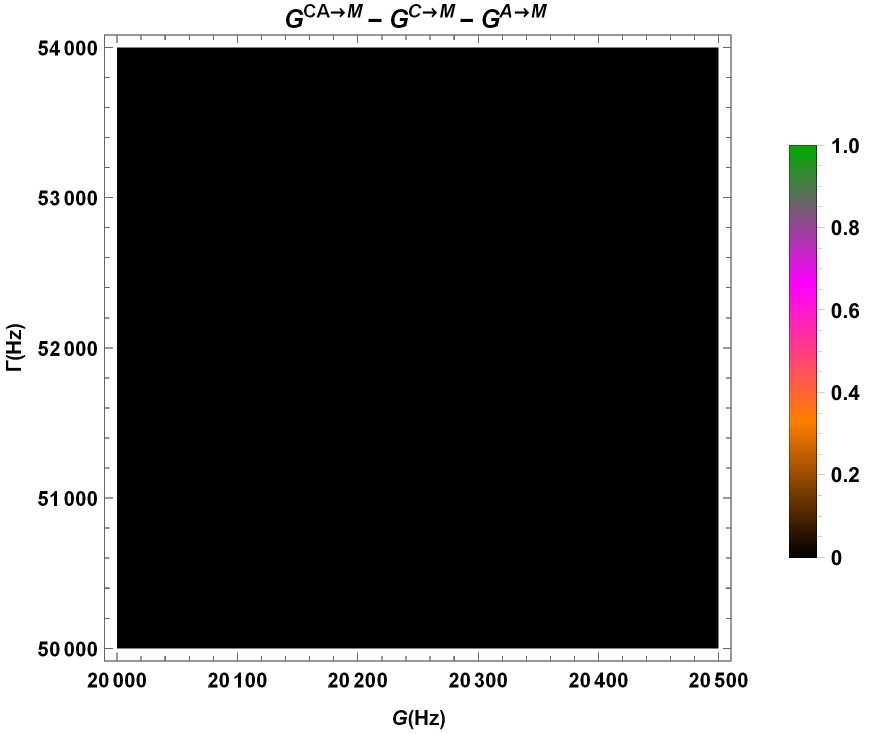}}
\caption{The value of the quantity $G^{CA \rightarrow M} - G^{C \rightarrow M} - G^{A \rightarrow M}$ remains zero.}\label{}
\end{subfigure}
\caption{The Gaussian steering $G^{C\rightarrow M} (a)$, $ G^{A\rightarrow M}(b)$ and $ G^{CA\rightarrow M}(c) $ behaviors versus $ \Gamma $ and $G $; $(d)$: Effect of $ \Gamma $ and $G $ on steering monogamy $G^{CA \rightarrow M} - G^{C \rightarrow M} - G^{A \rightarrow M}$, for $L= 1$ mm, $ \gamma_{A}=2\pi\times10^{3}$Hz, $k=2\pi\times10^{3}$Hz and $T=1\mu$k.}\label{}
\end{figure}
\begin{figure}[H]
\begin{subfigure}[b]{.48\linewidth}
\fbox{\includegraphics[width=\linewidth]{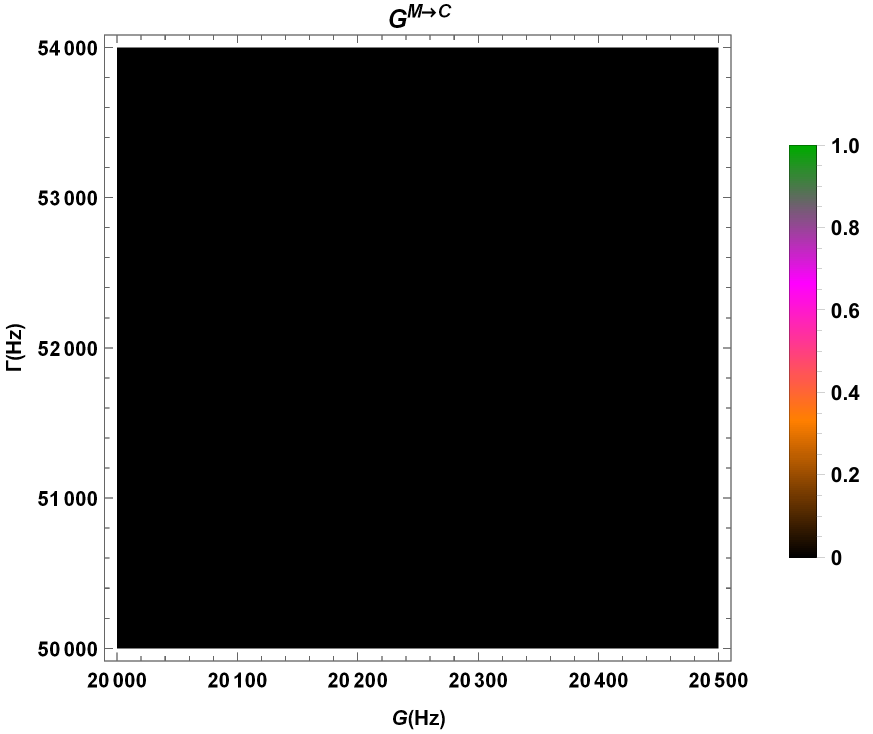}}
\caption{The steering of the mode $(C)$ by the mode $(M)$ is not authorized.}\label{}
\end{subfigure}
\begin{subfigure}[b]{.48\linewidth}
\fbox{\includegraphics[width=\linewidth]{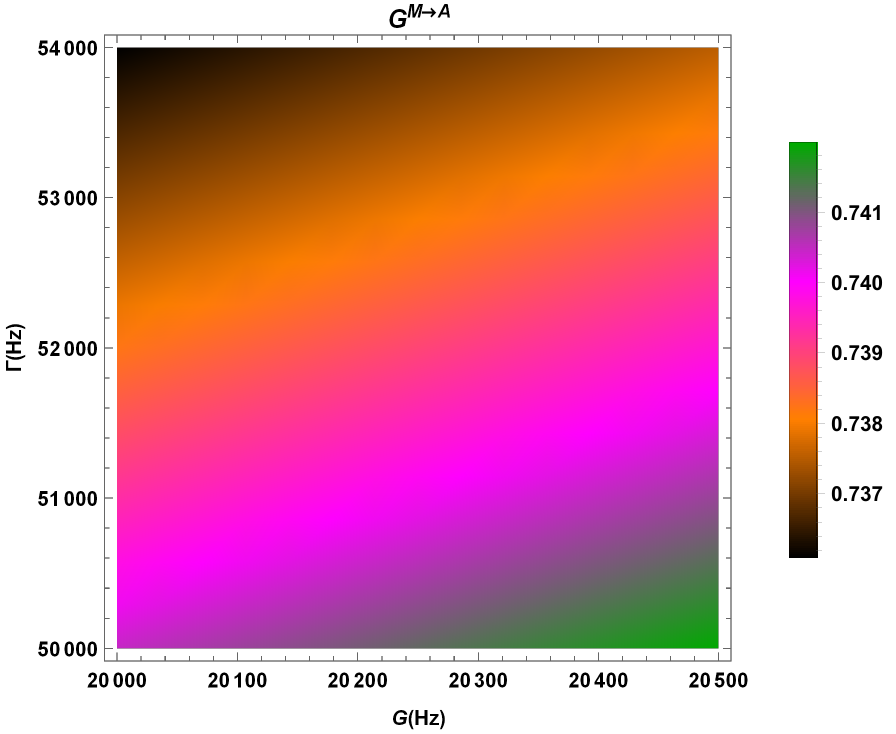}}
\caption{For higher value of $ \Gamma $ and a lowest value of $G$,  $G^{M \rightarrow A}$ is zero.}\label{}
\end{subfigure}
\begin{subfigure}[b]{.48\linewidth}
\fbox{\includegraphics[width=\linewidth]{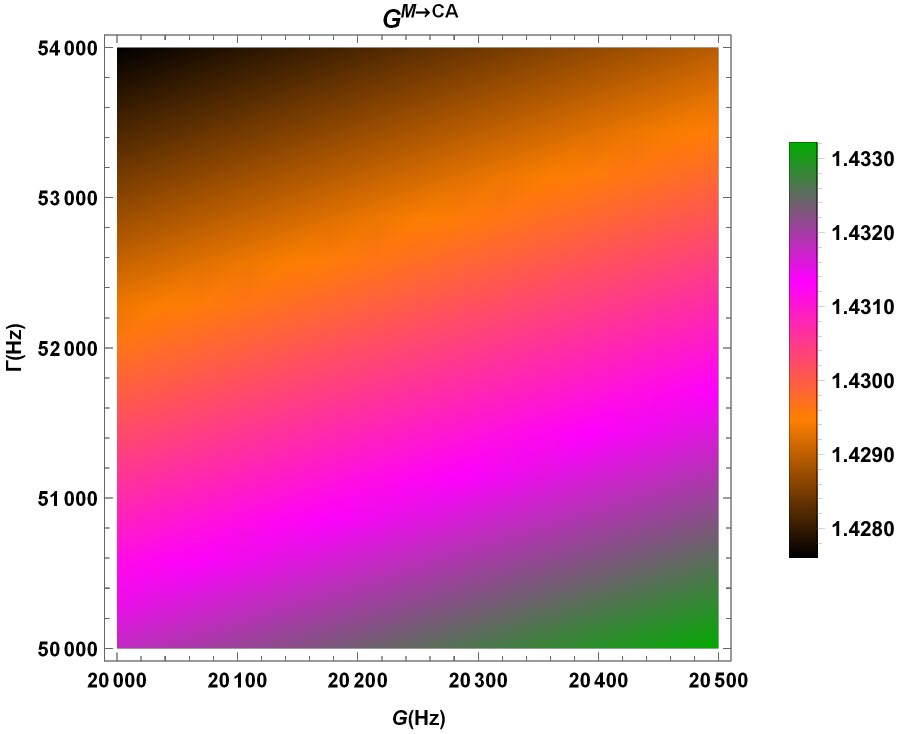}}
\caption{$G^{M\rightarrow CA}$ has the same behavior as that of $ G^{M \rightarrow A} $.}\label{}
\end{subfigure}
\hspace{0.5cm}
\begin{subfigure}[b]{.48\linewidth}
\fbox{\includegraphics[width=\linewidth]{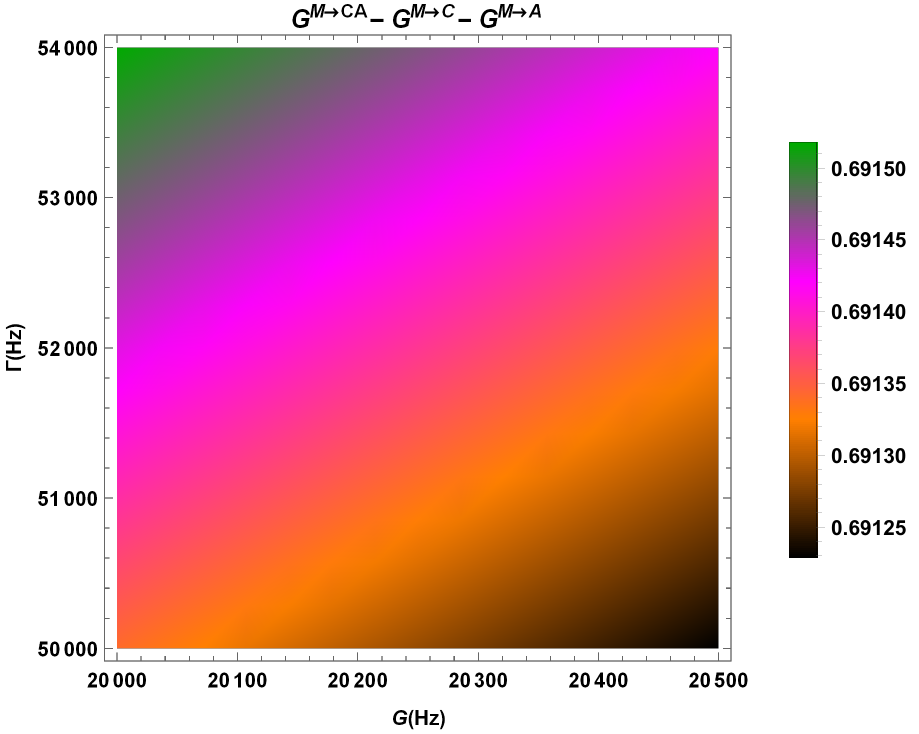}}
\caption{Steering monogamy inequality is fully satisfied.}\label{}
\end{subfigure}
\caption{The Gaussian steering $G^{M \rightarrow C} (a)$, $ G^{M \rightarrow A}(b)$ and $ G^{M \rightarrow CA}(c) $ behaviors versus $ \Gamma $ and $G $; $(d)$: Effect of $ \Gamma $ and $G $ on steering monogamy  $ G^{M \rightarrow CA} - G^{M \rightarrow C} - G^{M \rightarrow A}$, for $L= 1$ mm, $ \gamma_{A}=2\pi\times10^{3}$Hz, $k=2\pi\times10^{3}$Hz and $T=1 \mu$k.}\label{}
\end{figure}
The results presented in Fig. $5$, indicate that despite the variations of the parameters $ \Gamma $ and $ G $, we always have $G^{C \rightarrow M }=G^{A \rightarrow M }=G^{CA \rightarrow M }=0$, which shows that the mode $ M $ cannot be driven by the other remaining modes $A$ and $C$ neither collectively nor individually. Thus, the value of $G^{CA \rightarrow M } - G^{A \rightarrow M } - G^{C \rightarrow M }$ remains zero, this means that the parameters $ G $ and $ \Gamma $ have no effect on the evolution of steering monogamy inequality.\\
As shown in Fig. $6$, the degree of steering among the three modes $M$, $A $ and $C$ is governed by the CKW-type steering monogamy relations, i.e,  $G^{M \rightarrow CA } - G^{M \rightarrow C} - G^{M \rightarrow A } \geq 0$. Moreover, unlike to the variation of $G^{M \rightarrow CA }\geq 0$ and $G^{M \rightarrow A }\geq 0$  amounts, the quantity $G^{M \rightarrow CA } - G^{M \rightarrow C  } - G^{M \rightarrow A }$ shows a slight rise by increasing the $\Gamma$ values and decreasing the values of $G$. Based on Figs. $5$ and $6$, the studied tripartite state is steerable only from $M \rightarrow CA$, which proving a one-way steering class. For a higher values of $G$ with a lesser values of $\Gamma$, the steering is not authorized in any direction. 
\begin{figure}[H]
\begin{subfigure}[b]{.48\linewidth}
\fbox{\includegraphics[width=\linewidth]{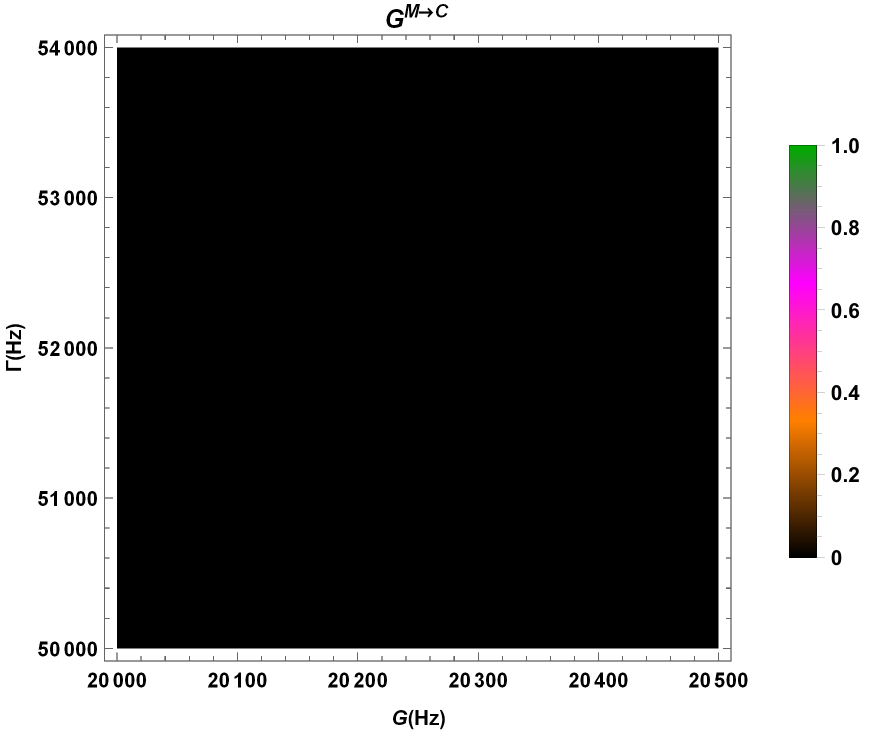}}
\caption{The steering of the mode $(C)$ by the mode $(M)$ is not authorized}\label{}
\end{subfigure}
\begin{subfigure}[b]{.48\linewidth}
\fbox{\includegraphics[width=\linewidth]{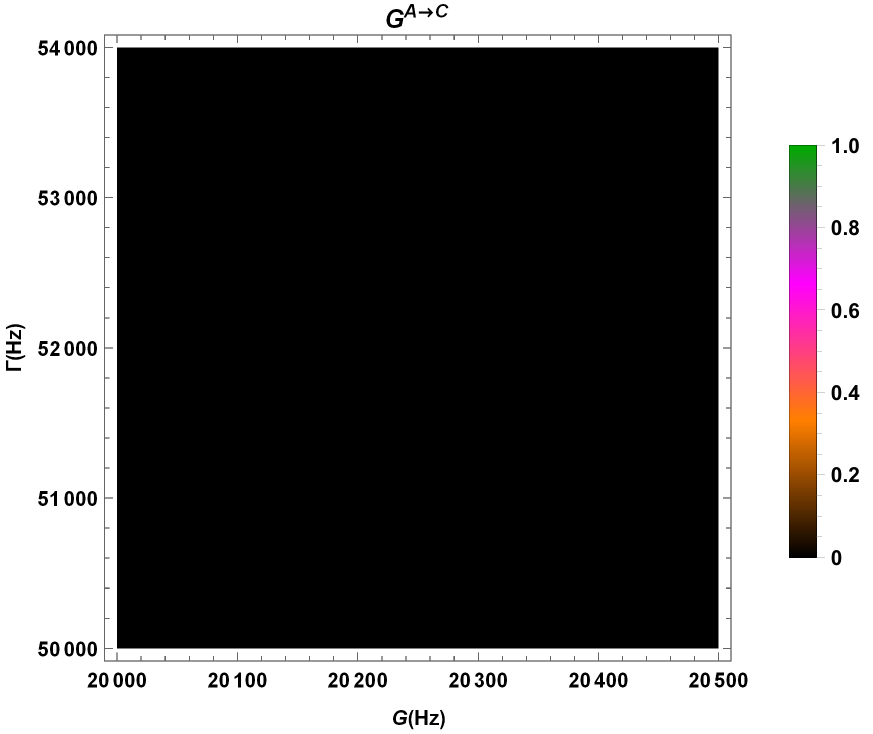}}
\caption{The steering of the mode $(C)$ by the mode $(A)$ is not authorized}\label{}
\end{subfigure}
\begin{subfigure}[b]{.48\linewidth}
\fbox{\includegraphics[width=\linewidth]{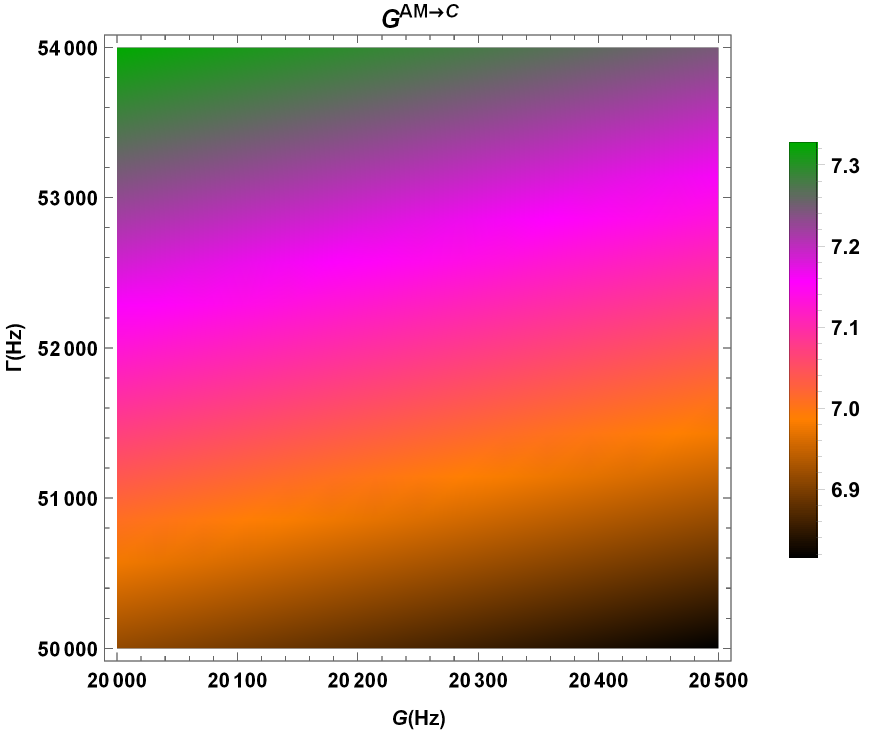}}
\caption{For higher value of $ G$ and lowest value of $\Gamma$,  $ G^{AM \setminus C} $ is zero.}\label{}
\end{subfigure}
\hspace{0.5cm}
\begin{subfigure}[b]{.48\linewidth}
\fbox{\includegraphics[width=\linewidth]{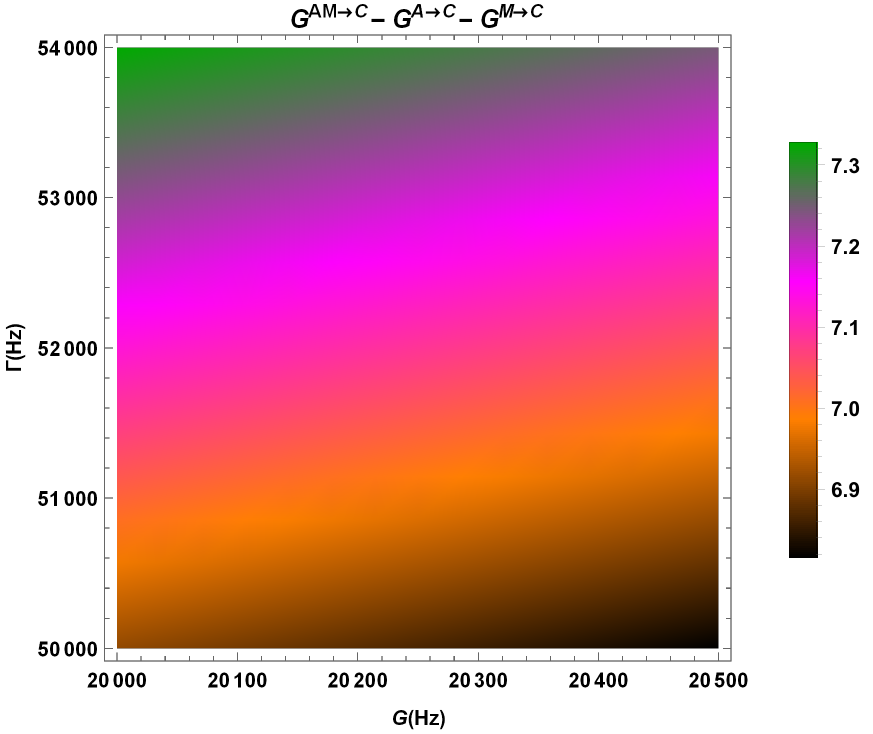}}
\caption{Steering monogamy inequality is fully satisfied, i.e,  $G^{AM \rightarrow C } - G^{A \rightarrow C } - G^{M \rightarrow C }\geq 0$.}\label{}
\end{subfigure}
\caption{The Gaussian steering $G^{M \rightarrow C} (a)$, $ G^{A \rightarrow C}(b)$ and $ G^{AM \rightarrow C}(c) $ behaviors versus $ \Gamma $ and $G $; $(d)$: Effect of $ \Gamma $ and $G $ on steering monogamy  $ G^{AM \rightarrow C} - G^{A \rightarrow C} - G^{M \rightarrow C} $, for $L= 1$ mm, $ \gamma_{A}=2\pi\times10^{3}$Hz, $k=2\pi\times10^{3}$Hz and $T=1\mu$k.}\label{}
\end{figure}

\begin{figure}[H]
\begin{subfigure}[]{.48\linewidth}
\fbox{\includegraphics[width=\linewidth]{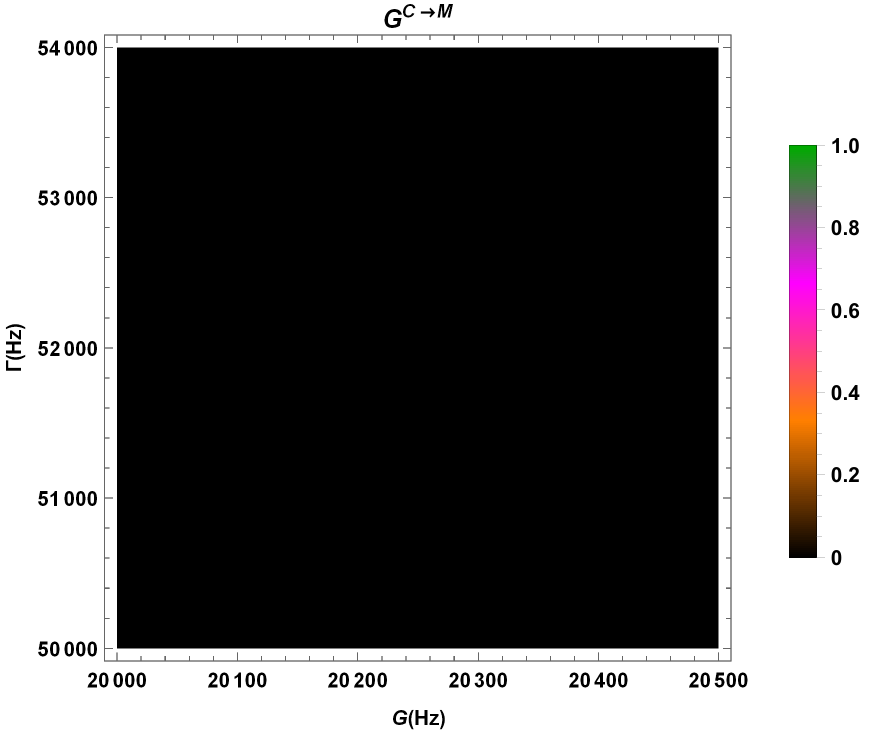}}
\caption{The steering of the mode $(M)$ by the mode $(C)$ is not authorized.}\label{}
\end{subfigure}
\begin{subfigure}[]{.48\linewidth}
\fbox{\includegraphics[width=\linewidth]{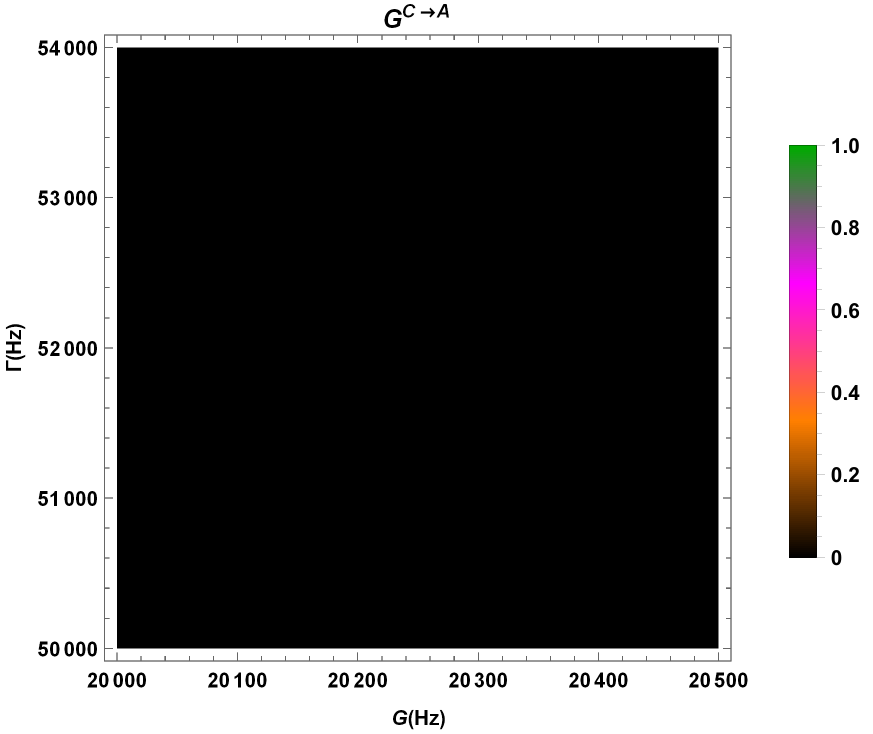}}
\caption{The steering of the mode $(A)$ by the mode $(C)$ is not authorized.}\label{}
\end{subfigure}
\begin{subfigure}[]{.48\linewidth}
\fbox{\includegraphics[width=\linewidth]{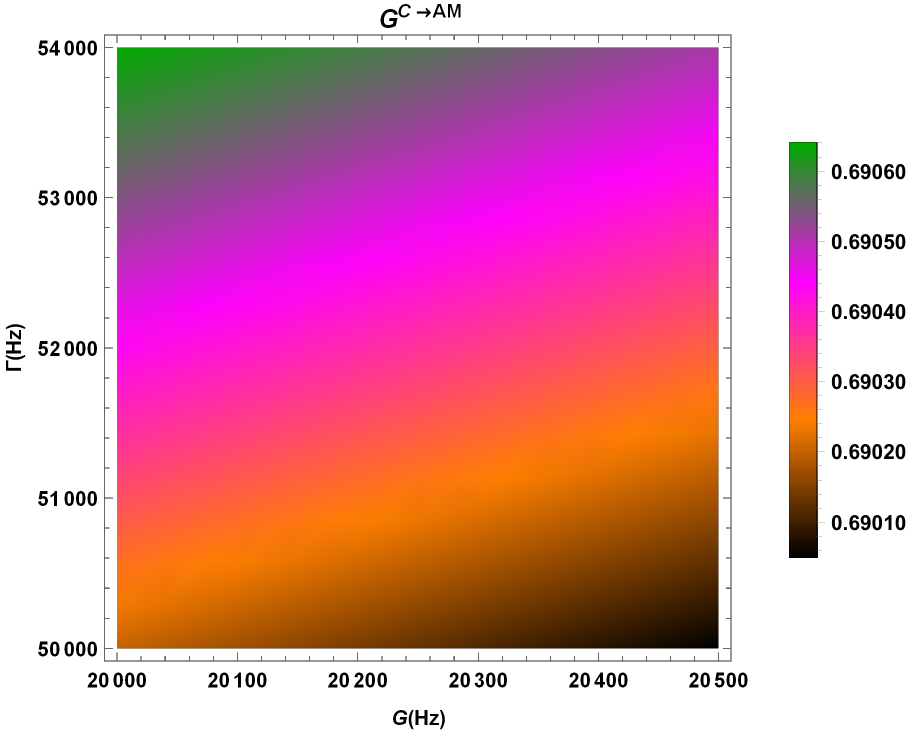}}
\caption{For higher value of $ G$ and lowest value of $\Gamma$,  $ G^{C \rightarrow AM} $ is zero.}\label{}
\end{subfigure}
\hspace{0.5cm}
\begin{subfigure}[]{.48\linewidth}
\fbox{\includegraphics[width=\linewidth]{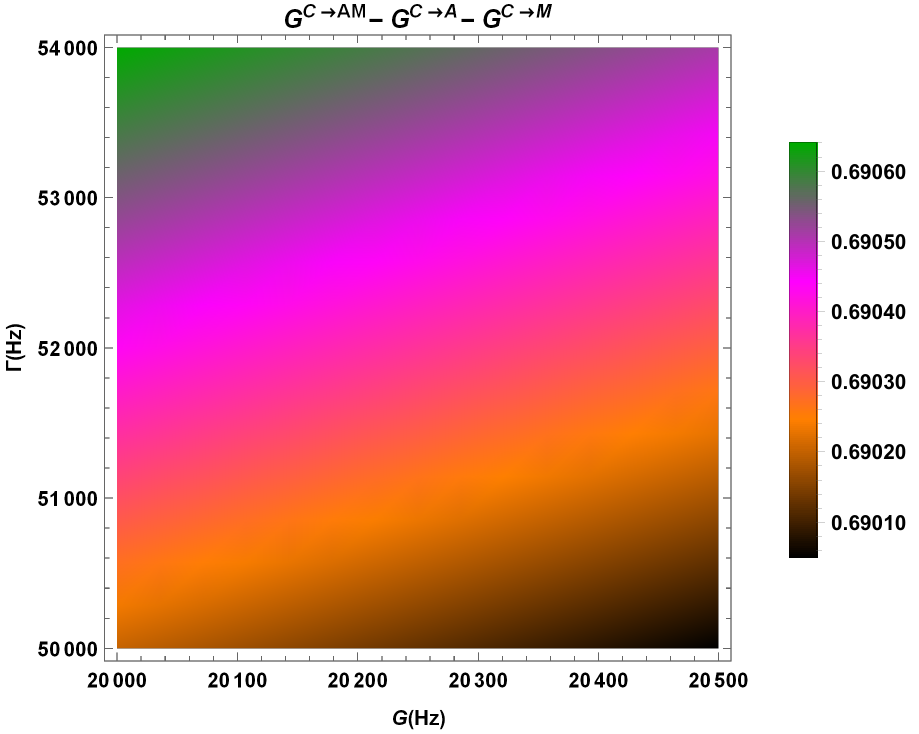}}
\caption{Steering monogamy inequality is fully satisfied, i.e,  $G^{C \rightarrow AM} - G^{C \rightarrow A} - G^{C \rightarrow M} 
 \geq 0$.}\label{}
\end{subfigure}
\caption{The Gaussian steering $G^{C \rightarrow M} (a)$, $ G^{C \rightarrow A}(b)$ and $ G^{C \rightarrow AM}(c) $ behaviors versus $ \Gamma $ and $G $; $(d)$: Effect of $ \Gamma $ and $G $ on steering monogamy  $ G^{C \rightarrow AM} - G^{C \rightarrow A} - G^{C \rightarrow M}$, for $L= 1$ mm, $ \gamma_{A}=2\pi\times10^{3}$Hz, $k=2\pi\times10^{3}$Hz and $T=1\mu$k.}
\end{figure}
In the absence of bimodal steering, i.e, $ \lbrace G^{M \rightarrow C }=G^{A \rightarrow C }=G^{C \rightarrow M }=G^{C \rightarrow A} = 0\rbrace $, we observe a strong collective steering of mode $C$ by modes $A$ and $M$ ($ G^{AM \rightarrow C}$) comparatively to the collective steering of modes $A$ 
and $M$ by mode $C$ ($ G^{C \rightarrow AM}$). The quantities $ G^{AM \rightarrow C } - G^{A \rightarrow C } - G^{M \rightarrow C } $ and $ G^{C \rightarrow AM } - G^{C \rightarrow A } - G^{C \rightarrow M }$ have similar behaviors, i.e, if we contain to increase the $G$ values and decrease those of $\Gamma$  values, the quantities $ G^{AM \rightarrow C } - G^{A \rightarrow C } - G^{M \rightarrow C }   $ and $ G^{C \rightarrow AM } - G^{C \rightarrow A } - G^{C \rightarrow M }$ become zero.
In this limit, no steering can exist between modes. Steering monogamy inequalities  $\lbrace G^{AM \rightarrow C } -  G^{A \rightarrow C } -  G^{M \rightarrow C }\geq 0 $ and $ G^{C \rightarrow AM } - G^{C \rightarrow A } - G^{C \rightarrow M }  \geq 0 \rbrace $ are fully satisfied, and the holding of the two inequalities $  G^{AM \rightarrow C } - G^{A \rightarrow C } - G^{M \rightarrow C } > 0 $ and $ G^{C \rightarrow AM } - G^{C \rightarrow A } - G^{C \rightarrow M }  > 0 $ in the same region, e.g, for $ 50 \times10^{3}$Hz$ \leq \Gamma \leq 54 \times10^{3}$Hz  and $20\times 10^{3}$Hz$ \leq G \leq 20.2 \times10^{3} $Hz, indicates  the existence of two-way steering, meaning that the tripartite state is steerable from $ C\rightarrow AM $ and $ AM\rightarrow C $. 
\begin{figure}[H]
\begin{subfigure}[b]{.48\linewidth}
\fbox{\includegraphics[width=8cm]{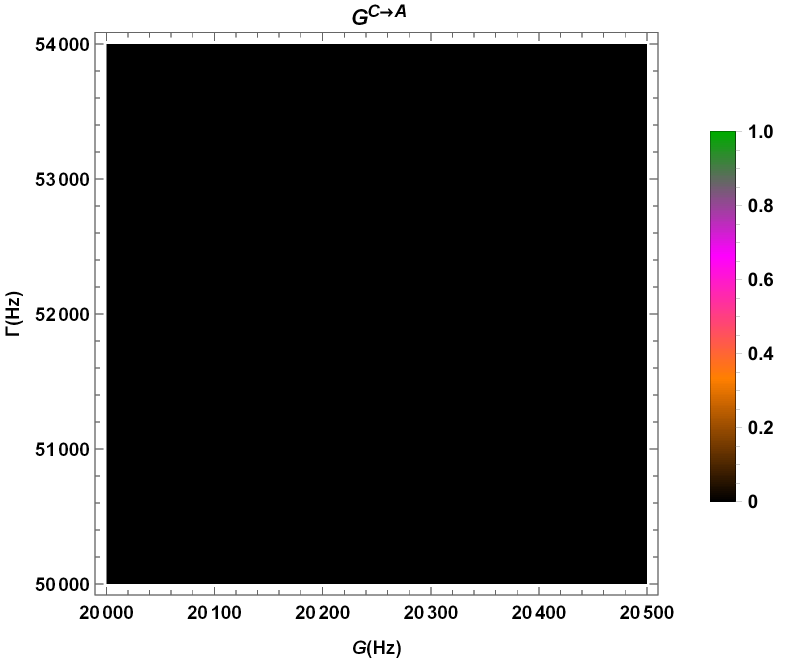}}
\caption{The steering of the mode $(A)$ by the mode $(C)$ is not authorized.}\label{}
\end{subfigure}
\begin{subfigure}[b]{.48\linewidth}
\fbox{\includegraphics[width=8cm]{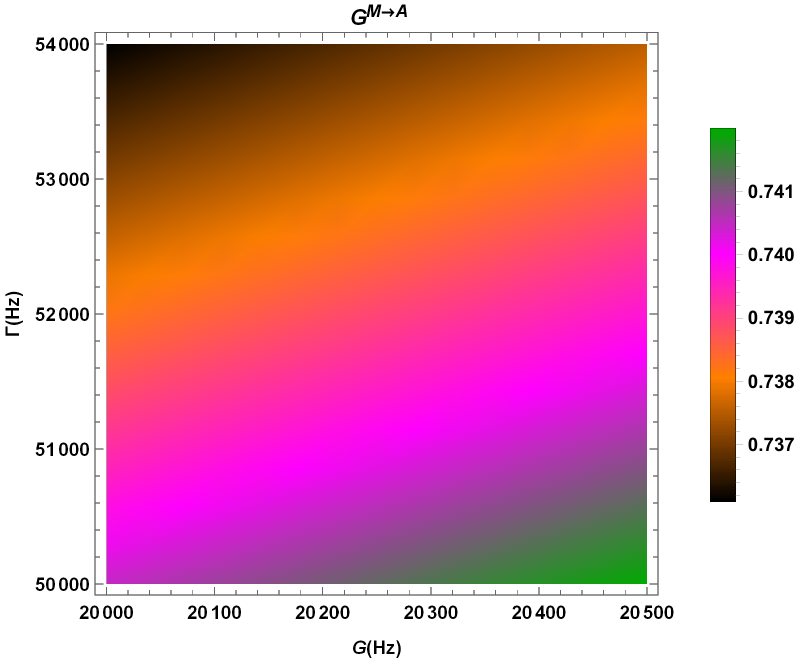}}
\caption{For higher value of $\Gamma$ and a lowest value of $G$,  $ G^{M\rightarrow A} $ is zero.}\label{}
\end{subfigure}
\begin{subfigure}[b]{.48\linewidth}
\fbox{\includegraphics[width=8cm]{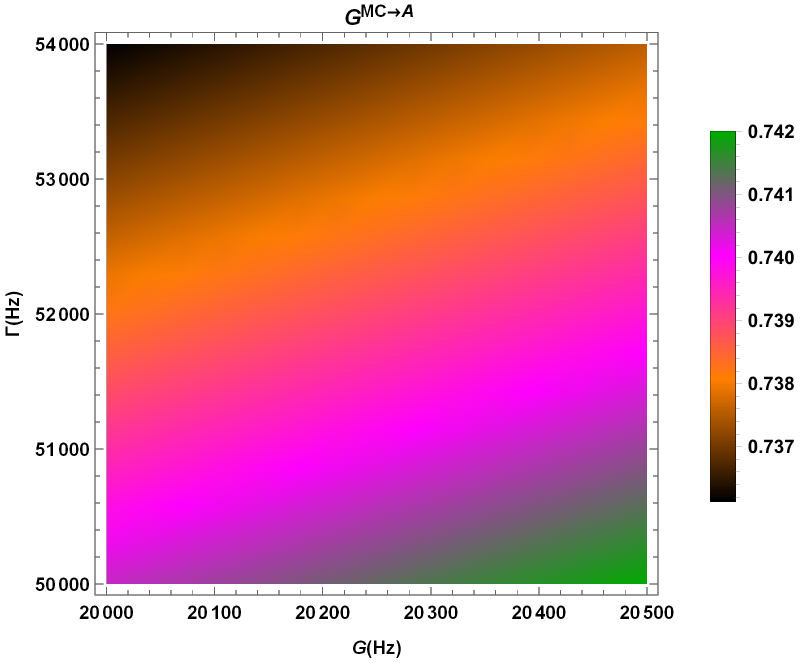}}
\caption{$ G^{MC\rightarrow A} $ increases by increasing the values of $G$ and decreasing the $\Gamma$ values.}\label{}
\end{subfigure}
\hspace{0.5cm}
\begin{subfigure}[b]{.48\linewidth}
\fbox{\includegraphics[width=8cm]{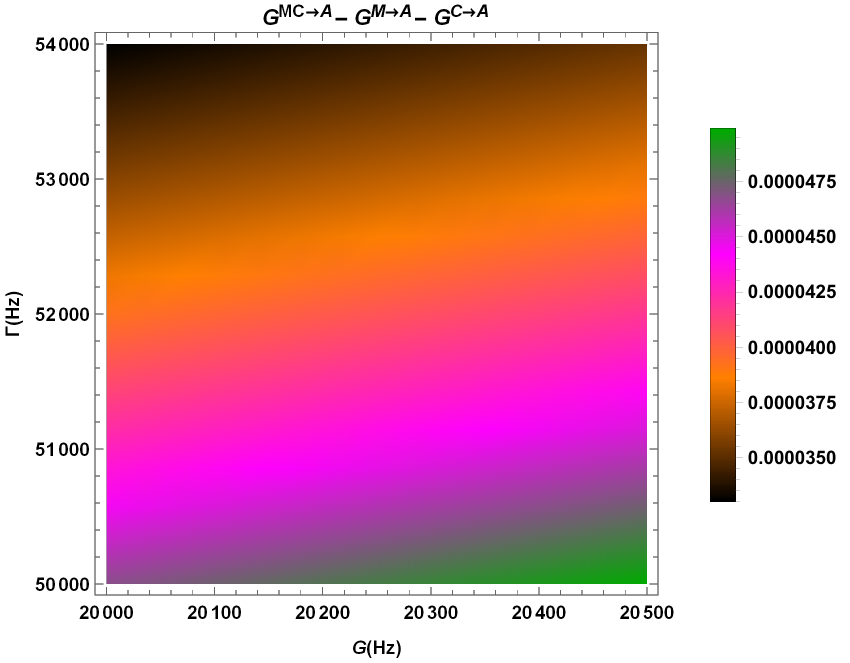}}
\caption{Steering monogamy inequality is fully satisfied, i.e,  $G^{MC \rightarrow A} - G^{M \rightarrow A} - G^{C \rightarrow A}   \geq 0$.}\label{}
\end{subfigure}
\caption{The Gaussian steering $G^{C \rightarrow A} (a)$, $ G^{M \rightarrow A}(b)$ and $ G^{MC \rightarrow A}(c) $ behaviors versus $ \Gamma $ and $G $; $(d)$: Effect of $ \Gamma $ and $G $ on steering monogamy  $ G^{MC \rightarrow A} - G^{M \rightarrow A} - G^{C \rightarrow A}  $, for $L= 1$ mm, $ \gamma_{A}=2\pi\times10^{3}$Hz, $k=2\pi\times10^{3}$Hz and $T=1\mu$k.}
\end{figure}

\begin{figure}[H]
\begin{subfigure}[b]{.48\linewidth}
\fbox{\includegraphics[width=\linewidth]{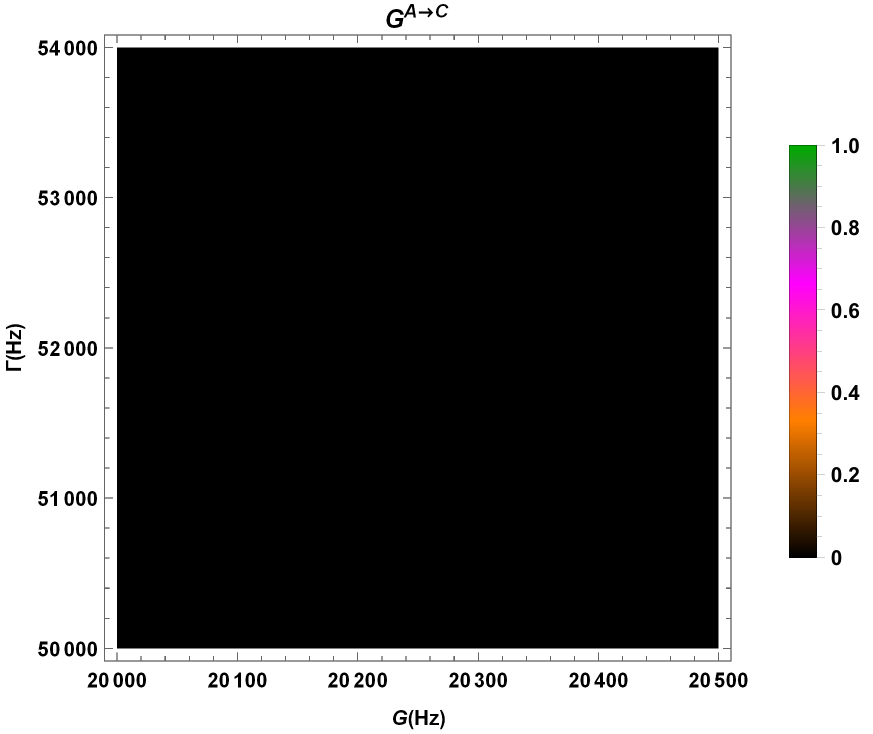}}
\caption{The steering of the mode $(C)$ by the mode $(A)$ is not authorized.}\label{}
\end{subfigure}
\begin{subfigure}[b]{.48\linewidth}
\fbox{\includegraphics[width=\linewidth]{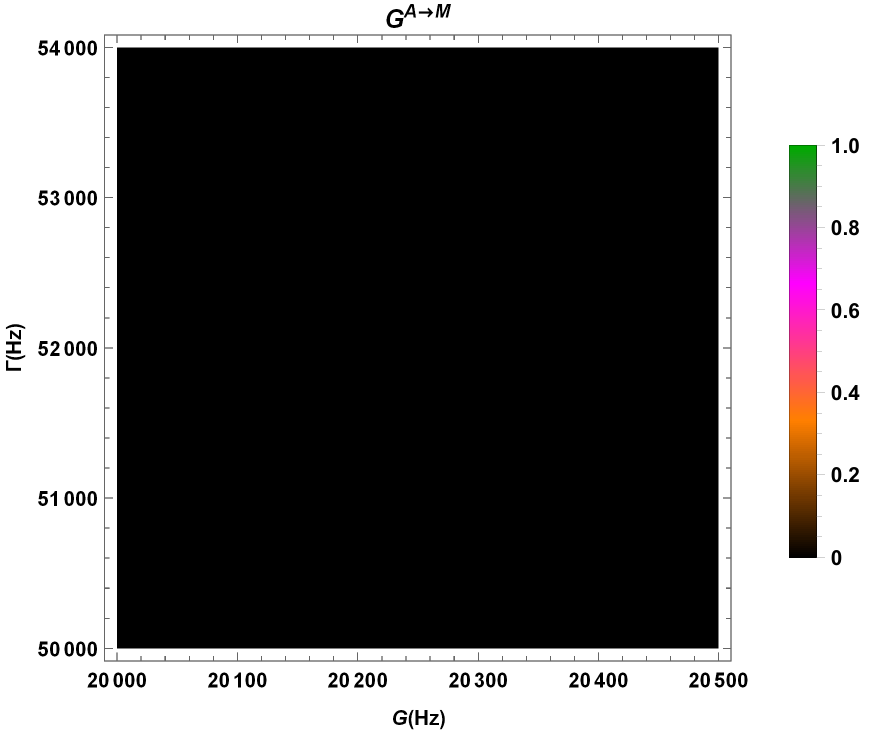}}
\caption{The steering of the mode $(M)$ by the mode $(A)$ is not authorized.}\label{}
\end{subfigure}
\begin{subfigure}[b]{.48\linewidth}
\fbox{\includegraphics[width=\linewidth]{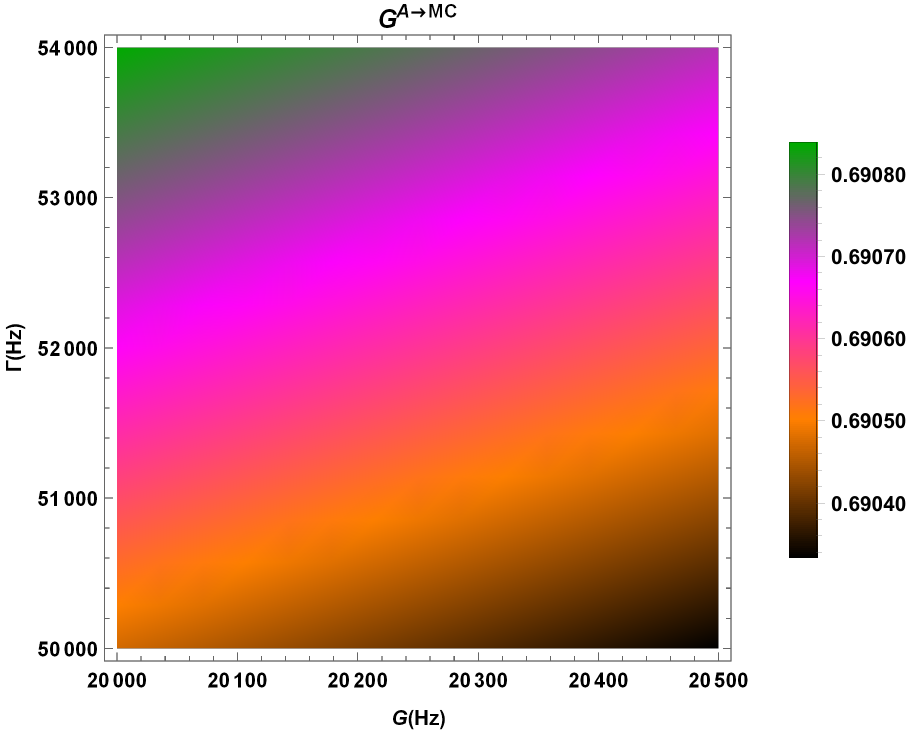}}
\caption{$G^{A \rightarrow MC}$ increases by increasing the values of $\Gamma$ and decreasing the $G$ values.}\label{}
\end{subfigure}
\hspace{0.5cm} 
\begin{subfigure}[b]{.48\linewidth}
\fbox{\includegraphics[width=\linewidth]{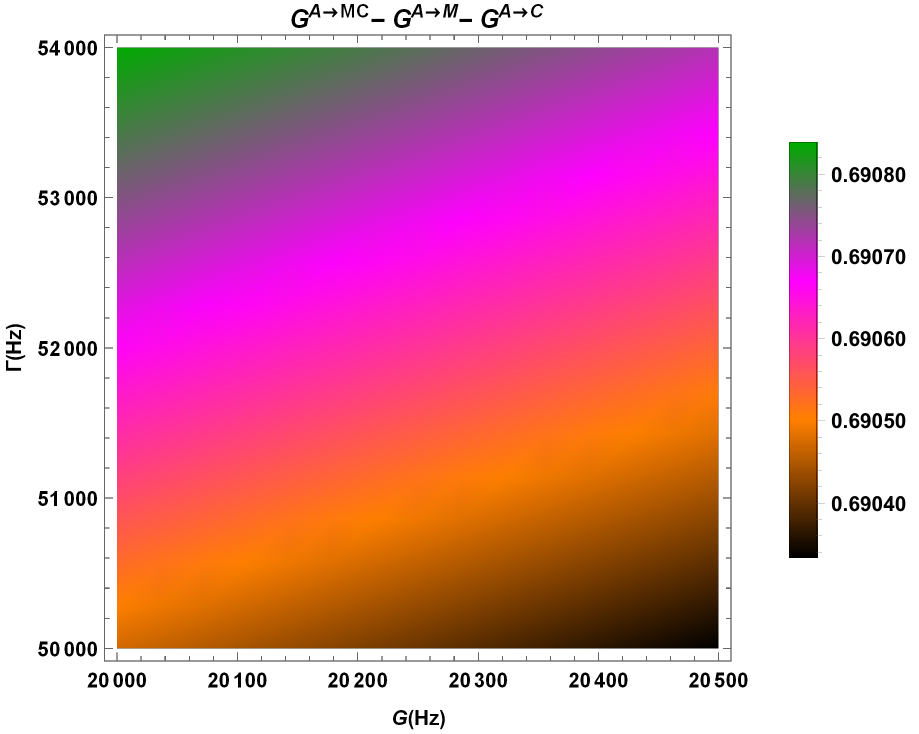}}
\caption{Steering monogamy inequality is fully satisfied, i.e, $G^{A \rightarrow MC} - G^{A \rightarrow M} - G^{A \rightarrow C} \geq 0$.
}\label{}
\end{subfigure}
\caption{The Gaussian steering $G^{A \rightarrow C} (a)$, $ G^{A \rightarrow M}(b)$ and $ G^{A \rightarrow MC}(c) $ behaviors versus $ \Gamma $ and $G $; $(d)$: Effect of $ \Gamma $ and $G $ on steering monogamy  $ G^{A \rightarrow MC} - G^{A \rightarrow M} - G^{A \rightarrow C}  $, for $L= 1$ mm, $ \gamma_{A}=2\pi\times10^{3}$Hz, $k=2\pi\times10^{3}$Hz and $T=1\mu$k.}\label{}
\end{figure}
Let us now examine the effects of the parameters $G$ and $\Gamma$ on the steering behaviors in the configurations $G^{C \rightarrow A} $, $ G^{M \rightarrow A}  $ and $G^{MC \rightarrow A} $, by analyzing the results presented in Fig. $9$.  We remarkably observe that $G^{MC \rightarrow A} - G^{M \rightarrow A} - G^{C \rightarrow A} $ behaves very similarly to $G^{MC \rightarrow A} $ and $G^{M \rightarrow A} $, so that the degree of steering remains robust to the increase in  $\Gamma$ values up to $\Gamma = 53.6\times10^{3} $Hz, unlike the $G^{A \rightarrow MC} - G^{A \rightarrow M} - G^{A \rightarrow C} $ and $G^{A \rightarrow MC} $ behaviors (Fig. $10$). In addition, we notice that the inequalities $G^{MC \rightarrow A} - G^{M \rightarrow A} - G^{C \rightarrow A} \geq 0 $ and  $G^{A \rightarrow MC} - G^{A \rightarrow M} - G^{A \rightarrow C}  \geq 0 $ are fully satisfied. Moreover, if we choose any value of $\Gamma$ and $G$, e.g, in the ranges $ 51 \times10^{3} $Hz$ \leq \Gamma \leq 53.6 \times10^{3}$Hz  and $20\times 10^{3}$Hz$ \leq G \leq 20.3 \times10^{3}  $Hz respectively, the inequalities $G^{MC \rightarrow A} - G^{M \rightarrow A} - G^{C \rightarrow A} > 0  $ and $G^{A \rightarrow MC} - G^{A \rightarrow M} - G^{A \rightarrow C} > 0  $ are maintained simultaneously, which designates the existence of two-way steerability. 
To conclude this paragraph, we can therefore say that one-way steering can be obtained either by taking the highest possible value of $\Gamma$  [approximately  $ 50.4 \times 10^{3}$Hz] with the lowest value of $G$ or by decreasing $\Gamma$  until its lowest value and simultaneously increasing $G$ up to the maximum value, i.e, from $ 20.3 \times 10^{3}$Hz. 
\section{Conclusion}\label{7}
In conclusion, we have considered an atomic optomechanical system, we were interested in the evolution of the tripartite entanglement as well as that of the degree of Gaussian steering as a function of the cavity-collective atomic coupling ($\Gamma $) and effective optomechanical coupling ($G$). The behavior of tripartite entanglement based on CKW-type monogamy entanglement under all permutations of modes (mechanical mode $M$, optical mode $C$ and atomic mode $A$), was studied in a stable region defined by ($ 50 \times10^{3} $Hz$ \leq \Gamma \leq 54 \times10^{3}$Hz and $20\times 10^{3}$Hz$ \leq G \leq 20.5 \times10^{3} $Hz), exploiting the evolution of pairwise entanglement under the same conditions. As a result, the behavior of the tripartite entanglement in the configurations $ \varepsilon^{A\setminus MC} $, $ \varepsilon^{A\setminus M} $ and $ \varepsilon^{A \setminus C} $ is similar of that of tripartite entanglement in configurations $ \varepsilon^{M\setminus CA} $, $ \varepsilon^{M\setminus C} $ and $ \varepsilon^{M\setminus A} $. More interesting, the CKW-type  entanglement monogamy inequality is satisfy under all permutations of different modes. The considered three-mode Gaussian state shows genuine tripartite entanglement quantified by $ \varepsilon^{A\setminus MC} - \varepsilon^{A\setminus M}-\varepsilon^{A\setminus C} $, which can be improved by decreasing the collective atomic-cavity coupling $\Gamma$ for sufficiently large value of $G$. Tripartite steering was also been studied, and it is interesting to note that the studied three-mode Gaussian state may exhibit, depending on a specific conditions, either a two-way steering or a one-way steering.

\end{document}